

\magnification 1100
\def\m@th{\mathsurround=0pt}
\newif\ifdtpt
\def\displ@y{\openup1\jot\m@th
    \everycr{\noalign{\ifdtpt\dt@pfalse
    \vskip-\lineskiplimit \vskip\normallineskiplimit
    \else \penalty\interdisplaylinepenalty \fi}}}
\def\eqalignc#1{\,\vcenter{\openup1\jot\m@th
                \ialign{\strut\hfil$\displaystyle{##}$\hfil&
                              \hfil$\displaystyle{{}##}$\hfil&
                              \hfil$\displaystyle{{}##}$\hfil&
                              \hfil$\displaystyle{{}##}$\hfil&
                              \hfil$\displaystyle{{}##}$\hfil\crcr#1\crcr}}\,}
\def\eqalignnoc#1{\displ@y\tabskip\centering \halign to \displaywidth{
                  \hfil$\displaystyle{##}$\hfil\tabskip=0pt &
                  \hfil$\displaystyle{{}##}$\hfil\tabskip=0pt &
                  \hfil$\displaystyle{{}##}$\hfil\tabskip=0pt &
                  \hfil$\displaystyle{{}##}$\hfil\tabskip=0pt &
                  \hfil$\displaystyle{{}##}$\hfil\tabskip\centering &
                  \llap{$##$}\tabskip=0pt \crcr#1\crcr}}
\def\leqalignnoc#1{\displ@y\tabskip\centering \halign to \displaywidth{
                  \hfil$\displaystyle{##}$\hfil\tabskip=0pt &
                  \hfil$\displaystyle{{}##}$\hfil\tabskip=0pt &
                  \hfil$\displaystyle{{}##}$\hfil\tabskip=0pt &
                  \hfil$\displaystyle{{}##}$\hfil\tabskip=0pt &
                  \hfil$\displaystyle{{}##}$\hfil\tabskip\centering &
                  \kern-\displaywidth\rlap{$##$}\tabskip=\displaywidth
                  \crcr#1\crcr}}
%

%


\def\charlvmidlw#1#2{\,\vtop{\ialign{##\crcr
      #1\crcr\noalign{\kern1pt\nointerlineskip}
      $\hfil#2\hfil$\crcr}}\,}
\def\charlvlowlw#1#2{\,\vtop{\ialign{##\crcr
      $\hfil#1\hfil$\crcr\noalign{\kern1pt\nointerlineskip}
      #2\crcr}}\,}
\def\charlvmidup#1#2{\,\vbox{\ialign{##\crcr
      $\hfil#1\hfil$\crcr\noalign{\kern1pt\nointerlineskip}
      #2\crcr}}\,}
\def\charlvupup#1#2{\,\vbox{\ialign{##\crcr
      #1\crcr\noalign{\kern1pt\nointerlineskip}
      $\hfil#2\hfil$\crcr}}\,}
\def\vspce{\kern4pt} \def\hspce{\kern4pt}    

\def\emptybox{\vbox{\kern.7ex\hbox{\kern.5em}\kern.7ex}}
 \font\sevmi  = cmmi7              
    \skewchar\sevmi ='177
 \font\fivmi  = cmmi5              
    \skewchar\fivmi ='177
\font\tenmib=cmmib10
\newfam\bfmitfam

\textfont\bfmitfam=\tenmib
\scriptfont\bfmitfam=\sevmi
\scriptscriptfont\bfmitfam=\fivmi
%

%
\def\mathcedilla{\vtop{\hbox{c}{\kern0pt\nointerlineskip}
	         {\hbox{$\mkern-2mu \mathchar"0018\mkern-2mu$}}}}

\mathchardef\gq="0060
\mathchardef\dq="0027
\mathchardef\ssmath="19
\mathchardef\aemath="1A
\mathchardef\oemath="1B
\mathchardef\omath="1C
\mathchardef\AEmath="1D
\mathchardef\OEmath="1E
\mathchardef\Omath="1F
\mathchardef\imath="10
\mathchardef\fmath="0166
\mathchardef\gmath="0167
\mathchardef\vmath="0176

%

\def\twodot{.\kern-0.1em.}

\def\paral{\mathrel{/\kern-.25em/}}
\def\grlo{\mathrel{\hbox{\lower.2ex\hbox{\rlap{$>$}\raise1ex\hbox{$<$}}}}}
\def\logr{\mathrel{\hbox{\lower.2ex\hbox{\rlap{$<$}\raise1ex\hbox{$>$}}}}}
\def\greq{\mathrel{\hbox{\lower1ex\hbox{\rlap{$=$}\raise1.2ex\hbox{$>$}}}}}
\def\loeq{\mathrel{\hbox{\lower1ex\hbox{\rlap{$=$}\raise1.2ex\hbox{$<$}}}}}
\def\grsim{\mathrel{\hbox{\lower1ex\hbox{\rlap{$\sim$}\raise1ex\hbox{$>$}}}}}
\def\losim{\mathrel{\hbox{\lower1ex\hbox{\rlap{$\sim$}\raise1ex\hbox{$<$}}}}}
%
\font\ninerm=cmr9
\def\uniset{\rlap{\ninerm 1}\kern.15em 1}

\def\emptysq{\mathbin{\vbox{\hrule\hbox{\vrule height1ex \kern.5em
                            \vrule height1ex}\hrule}}}
\def\emptyrect{\mathbin{\vbox{\hrule\hbox{\vrule height1ex \kern1em
                              \vrule height1ex}\hrule}}}
\def\rightonleftarrow{\mathrel{\hbox{\raise.5ex\hbox{$\rightarrow$}\ignorespaces
                                   \lower.5ex\hbox{\llap{$\leftarrow$}}}}}
\def\leftonrightarrow{\mathrel{\hbox{\raise.5ex\hbox{$\leftarrow$}\ignorespaces
                                   \lower.5ex\hbox{\llap{$\rightarrow$}}}}}

\def\bkB{{\rm I\kern-.17em B}}
\def\bkC{{\rm \kern.24em
            \vrule width.05em height1.4ex depth-.05ex
            \kern-.26em C}}
\def\bkD{{\rm I\kern-.17em D}}
\def\bkE{{\rm I\kern-.17em E}}
\def\bkF{{\rm I\kern-.17em F}}
\def\bkG{{\rm \kern.24em
            \vrule width.05em height1.4ex depth-.05ex
            \kern-.26em G}}
\def\bkH{{\rm I\kern-.22em H}}
\def\bkI{{\rm I\kern-.22em I}}
\def\bkJ{{\rm \kern.19em
            \vrule width.02em height1.5ex depth0ex
            \kern-.20em J}}
\def\bkK{{\rm I\kern-.22em K}}
\def\bkL{{\rm I\kern-.17em L}}
\def\bkM{{\rm I\kern-.22em M}}
\def\bkN{{\rm I\kern-.20em N}}
\def\bkO{{\rm \kern.24em
            \vrule width.05em height1.4ex depth-.05ex
            \kern-.26em O}}
\def\bkP{{\rm I\kern-.17em P}}
\def\bkQ{{\rm \kern.24em
            \vrule width.05em height1.4ex depth-.05ex
            \kern-.26em Q}}
\def\bkR{{\rm I\kern-.17em R}}
\def\bkT{{\rm \kern.24em
            \vrule width.02em height1.5ex depth 0ex
            \kern-.27em T}}
\def\bkU{{\rm \kern.30em
            \vrule width.02em height1.47ex depth-.05ex
            \kern-.32em U}}
\def\bkZ{{\rm Z\kern-.32em Z}}
%

\input epsf
\rightline{SphT 94-153}
\vskip .2 cm
\rightline{DAMTP 95-03}
\overfullrule=0pt
\baselineskip=18pt

\centerline{{\bf Flavour Changing Neutral Current Effects}}
\vskip .4 cm
\centerline{\bf From}
\vskip .4 cm
\centerline{ \bf   Flavour Dependent Supergravity Couplings}
\vskip 2.2 cm
\centerline{\bf Ph. Brax}
\centerline{Department of Applied Mathematics and Theoretical Physics}
\centerline{Silver Street, Cambridge CB3 9EW, ENGLAND}
\vskip 12pt
\centerline{\bf C.A. Savoy}
\centerline{CEA, Service de Physique Th\'eorique, CE-Saclay}
\centerline{F-91191 Gif-sur-Yvette Cedex, FRANCE}
\vskip 2 cm
{\bf ABSTRACT}

We investigate the impact on low energy flavour mixing phenomena  of the
renormalisation group running of parameters in supersymmetric
theories. We have explicitly chosen
 flavour dependent supergravity couplings.
 The renormalisation group equations
are displayed in general matrix form. We work in a
scale dependent basis such that the hierarchy in the quark masses can be more
easily utilised. Simple analytic solutions are provided in matrix form (for
general boundary conditions) when only the third generation
 Yukawa couplings are retained in the renormalisation group equations.
We solve  the evolution equations of the KM matrix and the supersymmetry
breaking soft terms in the squark sector. We suggest a
general parametrisation of the soft analytic terms,  more suitable
for non-universal supergravity couplings. We point out that for flavour
dependent soft terms, there are new contributions to the neutron
electric dipole moment . Finally, we consider a model with maximal flavour
mixing in order
to estimate the effects due to the renormalisation scale dependence.
\vfill\eject
\noindent {\bf 1. INTRODUCTION}

\vskip 12pt

Twenty years ago, the discovery of the charmed quark established the GIM
mechanism and naturally explained the approximate flavour conservation
 in neutral
current (FCNC) processes $ [1]. $ It was then generalised to
include
the third family with the introduction of the Kobayashi-Maskawa matrix. This
settled the standard model framework enabling one
to calculate, up to non-perturbative
effects, flavour changing neutral current phenomena as radiative corrections.
Consequently, the comparison of theoretical expectations and  experimental
data provides valuable information on possible new physics beyond the
standard model. On top of the traditional set of parameters which characterise
FCNC in the $ K $-system,  more recent studies of  $ B $-meson decays
have considerably enlarged the FCNC phenomenology. A  rich literature $ [2-10]
$
is available
 about FCNC restrictions on supersymmetric extensions of the standard
model. Nevertheless, both the LEP (and Tevatron) constraints on supersymmetric
theories and some fresh insight on spontaneously broken supergravities from
superstrings have encouraged a recent revival of this subject $ [11-14]. $

The basic supersymmetry induced FCNC effects are produced by the
supersymmetric  analogues of the standard model loop diagrams for neutral
current processes, where gauge bosons are exchanged. Quarks and vector bosons
are then replaced by squarks and gauginos. It is convenient to separate such
graphs into two kinds: (a)\nobreak\ Loops that generate FCNC effects in the
standard
model with the $ W $-bosons replaced by charginos. (b)\nobreak\ Flavour
preserving loops
in the standard model with gluons, photons and $ Z $-bosons replaced by
gluinos
and neutralinos. If equal-charge quark and squark mass matrices are not
diagonal in the same basis, their couplings to neutral gauginos can induce
FCNC effects. Of course, the gluino exchange diagrams proportional to $
\alpha^ 2_{ {\rm strong}} $
may be particularly relevant. There are several sources of flavour mixing in
gaugino couplings that we now turn to discuss.

Within
 the general framework of supergravity $ [15], $ a theory is defined by the
gauge and matter superfields, by their couplings to supergravity encoded in
the K\"ahler potential, and by the generalised gauge and Yukawa couplings. The
low-energy theory is then fixed by the values of the auxiliary fields which
provide the spontaneous breaking of supersymmetry. Therefore, the parameters
in the low-energy theory with broken supersymmetry are to be
determined
  within
the framework of a larger theory explaining all interactions including
gravity. Superstrings $ [16] $ are the best prototypes, despite
being  incomplete. Nevertheless, recent studies
$ [17-19,12] $ of the field theoretical limit of superstrings have
given
 hints on
the
general form of the relevant parts of the K\"ahler potential as well as on
their interpretation in terms of stringy symmetries. A conspicuous result of
superstring model calculations is that the three families of quark
superfields may couple to supergravity according to quite different terms in
the K\"ahler potential. Therefore, the minimal hypothesis in phenomenological
analyses, alleging
that supergravity couplings are flavour independent is not quite
confirmed by these explorations within superstring theories. The
 dilaton superfield in these theories does have universal
supergravity couplings to quark superfields $ [8,20]. $ But the moduli fields
associated to the compact 6-dimensional manifold, needed to reduce the
space-time dimension to four,  have  model dependent
couplings to the quark superfields $ [17,18,12]. $ Thus, the K\"ahler
potential can
be different for each flavour. Moreover, gauge singlet fields have been
conjectured to explain the so-called textures in the Yukawa couplings $
[21,22]. $
They are assumed to get relatively large vacuum expectation values and could
introduce non-universality corrections to the supergravity couplings of quark
superfields.

At the level of the effective (renormalisable) theory, below the Planck scale,
the supersymmetry breaking effects reduce to gaugino masses and the soft
interactions in the scalar potential $ [23]. $ There appear hermitian $ {\rm
(mass)}^ {\rm 2} $
terms involving the scalar fields of each given $ {\rm SU(3)\times SU(2)\times
U(1)} $ quantum
numbers and their complex conjugates. The corresponding sub-matrix of the
scalar $ {\rm (mass)}^ {\rm 2} $ matrix depend on the K\"ahler potential and
on the supersymmetry
breaking auxiliary fields $ [24,15]. $ In particular, the universality or
flavour
independence hypothesis assumes equal masses for all squarks at the
unification. At lower energies, this conjectured universality is broken by
radiative corrections due to Yukawa interactions and acquire some calculable
flavour dependence. Moreover, analytic $ {\rm (mass)}^ {\rm 2} $ terms
involving scalar fields
of opposite $ {\rm SU(3)}_c {\rm \times U(1)}_{ {\rm em}} $ quantum numbers
are generated after the electro-weak
gauge symmetry breaking by the Higgs scalars. They are basically proportional
to the Yukawa couplings in the superpotential. Again, if  universality is
assumed for the proportionality factors, referred to as $ A $-parameters, at
the
unification scale, their equality is spoilt at lower energies by the
calculable radiative corrections.

In most  discussions of supersymmetric FCNC processes,  universality
of soft terms is assumed. Then, the most striking effects of
 radiative corrections are of two kinds: a)\nobreak\ gauge corrections
which are
universal and, in general, tend to attenuate loop effects by an overall rise
in the squark masses if gauginos are relatively heavy, and b)\nobreak\ Yukawa
corrections dominated by the top coupling, $ \lambda_ t, $ tend to align the
down squark
mass eigenstates to the up quarks $ [3-5,7]. $ This reverses the pattern of
gaugino couplings in comparison with the gauge boson ones. Chargino couplings
to down squark and up quarks are approximately family diagonal while gluino
and neutralino couplings become proportional to the Kobayashi-Maskawa matrix.
However, the expected physical effects are generally quite tiny with the
present overall bounds on supersymmetric particles\footnote{$ ^1 $}{For a
recent
discussion, see e.g., Refs.$ [10,11]. $}. The $ b \longrightarrow s\gamma $
transition is the remarkable
exception which gives interesting information on supersymmetric theories $
[7,8,10]. $

Motivated by superstrings, as well as symmetries proposed to explain the
structure of Yukawa couplings, new analyses $[11-14] $ have been performed
on FCNC transitions produced by non-universality in supergravity
couplings.
If the flavour dependence is large already at the supergravity or Planck
scale, the quark mass eigenstates and the squark mass eigenstates are not
superpartners  under the action of supersymmetry. There is a
unitary transformation in the family space reflecting the difference of the
basis for physical quark and squark states. The gluino and neutralino
couplings will be proportional to this unitary matrix. The natural
supergravity basis for quarks and squarks is established by diagonalisation
of the K\"ahler metrics  (with its field-dependence classically fixed). In
this
supergravity basis, the squark $ {\rm (mass)}^ {\rm 2} $ matrix is dominantly
diagonal but the
eigenvalues may be different. In the same basis, the Yukawa couplings are
expected to be non-diagonal. In order to define the quark mass eigenstates,
one has to perform four unitary transformations to diagonalise the up and
down Yukawa couplings in the family space at the supergravity scale. These
unitary matrices tend to produce  FCNC effects in neutral
gaugino exchange already at tree level. But the difference of many orders
of magnitude between the unification scale and the low energy scale
suitable
for FCNC calculations, suggests that
the quantum running of the quark and squark masses and mixing matrices could
affect the tree-level pattern.

In this paper, we present a detailed study of the renormalisation group
equations for the Yukawa and soft-terms in the quark-squark sector, written
as general matrices in family space $ [4,5]. $ The phenomenological hierarchy
in
the Yukawa coupling eigenvalues is exploited using the obvious device of
keeping the up quark Yukawa couplings diagonal through scale dependent
rotations. The RGE for Yukawa couplings and soft terms are then easily solved
in the approximation in which all Yukawa couplings but $ \lambda_ t $ are
neglected. It
is straightforward to compute the corrections proportional to the other
couplings $ \left(\lambda^ 2_b,\lambda^ 2_c,\ {\rm etc} \ \ \ \right) $ to the
lowest order. Although they may be relevant
to some applications, these corrections are not included in this paper.

In Section 2, the RGE for the Kobayashi-Maskawa matrix is derived, then
solved in the case in which only the third family Yukawa couplings, $ \lambda_
t $ and $ \lambda_ b, $
are retained in the RGE. These results  already exist in the literature $
[4,25], $
but here, the derivation  is more transparent. In particular, the role of the
Yukawa coupling hierarchy is exhibited so that the accuracy of the
approximation can be readily estimated.

In Section 3, the RGE for the squark soft analytic terms
 are analysed. We argue that the current way
$ [15] $ of expressing their proportionality to the Yukawa couplings should be
changed to better implement the general boundary conditions at the
unification or Planck scales. Our parametrisation is more suitable because
Yukawa couplings relate  left-handed and right-handed quarks whose
supergravity properties may be very different. The RGE for the newly defined
matrices of parameters are displayed and analytically
 solved, in the limit of $
\lambda_ t $
dominance, in full complex matrix form.

In Section 4, the RGE for the hermitian sector of the $ {\rm (mass)}^ {\rm 2}
$ matrices are
analytically solved within the same approximation.

The calculation of the gluino loop contribution to the electric dipole moment
$ [26] $ of the neutron is revisited in Section 5, in view of possible effects
of
non-universality in the soft terms. We notice that contributions proportional
to $ m_b $ or $ m_t $ which are usually neglected, could overcome
current
contributions proportional to $ m_d $ or $ m_u. $ In order to estimate their
relative
strength, the upper limits $ [8,10,11] $ on the flavour mixing components of
the
scalar $ {\rm (mass)}^ {\rm 2} $ matrices are used.

In Section 6, we explore the effect of the RGE evolution on FCNC effects
in the presence of flavour dependent supergravity couplings. In the absence
of a realistic model, we define an ad hoc one, where the mixing matrices
relating the basis that diagonalise the K\"ahlerian metrics and the Yukawa
couplings, respectively, are maximal. With simple assumptions on the K\"ahler
potential and the superpotential, we find that in some extreme cases the
existent upper bounds on FCNC generating parameters can be saturated.
\vskip 17pt
\noindent {\bf 2. QUARK MIXING MATRICES}

\vskip 12pt

Let us begin with the RGE for Yukawas and KM matrices. We
define the unitary matrices $ U_R, $ $ U_L, $ $ V_R $ and $ V_L $ by the
diagonalisation of the
up and down Yukawa couplings:
$$ \eqalign{ U^{\dagger}_ R\lambda_ UU_L & = \left( \matrix{ \lambda_ u  &
&   \cr   &  \lambda_ c  &   \cr   &    &  \lambda_ t \cr} \right) \cr
V^{\dagger}_ R\lambda_ DV_L & = \left( \matrix{ \lambda_ d  &    &   \cr   &
\lambda_ s  &   \cr   &    &  \lambda_ b \cr} \right) \cr} \eqno (2.1) $$
where all the parameters have their dependence on $ t = {\rm ln} \left(\Lambda
/\Lambda_ 0 \right)/(4\pi)^ 2, $ defined
by the well-known RGE\footnote{$ ^2 $}{See, e.g., Ref.$ [4]. $}:
$$ \eqalign{{ {\rm d} \lambda_ U \over {\rm d} t} & = \lambda_ U \left[3
\left(\lambda^{ \dagger}_ U\lambda_ U+ {\rm Tr} \ \lambda^{ \dagger}_
U\lambda_ U \right)+\lambda^{ \dagger}_ D\lambda_ D-2C^u_\alpha g^2_\alpha
\right] \cr{ {\rm d} \lambda_ D \over {\rm d} t} & = \lambda_ D \left[3
\left(\lambda^{ \dagger}_ D\lambda_ D+ {\rm Tr} \ \lambda^{ \dagger}_
D\lambda_ D \right)+\lambda^{ \dagger}_ U\lambda_ U+ {\rm Tr} \ \lambda^{
\dagger}_ L\lambda_ L-2C^d_\alpha g^2_\alpha \right] \cr 2C^u_\alpha
g^2_\alpha &  = {16 \over 3} g^2_3 + 3g^2_2 + {13 \over 9} g^2_1 \cr
2C^d_\alpha g^2_\alpha &  = {16 \over 3} g^2_3 + 3g^2_\alpha  + {2 \over 9}
g^2_1 \cr} \eqno (2.2) $$

It is convenient to write $ \lambda_ U $ and $ \lambda_ D $ in terms of the
hermitian matrices
$$ Y_U = \lambda^{ \dagger}_ U\lambda_ U\ ,\ \ \ \ Y_D= \lambda^{ \dagger}_
D\lambda_ D \eqno (2.3) $$
diagonalised by the left-handed unitary transformations $ U_L $ and $ V_L $
only, and $ \lambda_ U\lambda^{ \dagger}_ U, $
$ \lambda_ D\lambda^{ \dagger}_ D $ diagonalised by $ U_R $ and $ V_R. $ We
then work with (2.3) only, the right-handed
counterpart being analogous. The RGE can then be cast in the form:
$$ \eqalign{{ {\rm d} Y_U \over {\rm d} t} & = \left\{ 3Y_U+Y_D,Y_U \right\}
+ 2\Delta_ UY_U \cr \Delta_ U & = 3\ {\rm Tr} \ Y_U - 2C^u_\alpha g^2_\alpha
\cr{ {\rm d} Y_D \over {\rm d} t} & = \left\{ 3Y_D+Y_U,U_D \right\} +2\Delta_
DY_D \cr \Delta_ D & = 3\ {\rm Tr} \ Y_D + {\rm Tr} \ Y _L-2C^D_\alpha
g^2_\alpha \cr} \eqno (2.4) $$

Now diagonalise $ Y_U $ and $ Y_D: $
$$ \eqalign{\hat Y_U & = U^{\dagger}_ LY_UU_L = \left( \matrix{ \lambda^ 2_u
&    &   \cr   &  \lambda^ 2_c  &   \cr   &    &  \lambda^ 2_t \cr} \right)
\cr\hat Y_D & = V^{\dagger}_ LY_DV_L = \left( \matrix{ \lambda^ 2_d  &    &
\cr   &  \lambda^ 2_s  &   \cr   &    &  \lambda^ 2_b \cr} \right) \cr} \eqno
(2.5) $$
and define the KM matrix (an analogous one may be defined for the
right-handed states)
$$ K_L = V^{\dagger}_ LU_L \eqno (2.6) $$

Then the RG evolution for the eigenvalues in $ \hat Y_U $ follows from (2.2):
$$ { {\rm d}\hat Y_U \over {\rm d} t} = \left[6\hat Y_U+2\Delta_ U + 2
\left(K^{\dagger}_ L\hat Y_DK_L \right)_{(d)} \right]Y_U \eqno (2.7) $$
where we denote the separation of any matrix into diagonal and non-diagonal
parts, $ M = M_{(d)} + M_{(nd)}. $
 Consider, e.g., $ U^{\dagger}_ L { {\rm d} \over {\rm d}
t} U_L, $ which is an element of $ {\rm SU(3)} /H $ where $ H $
is the $ {\rm SU(3)} $ Cartan algebra. The off-diagonal counterpart of (2.7)
coming from (2.2) reads:
$$ \left[U^{\dagger}_ L { {\rm d} U_L \over {\rm d} t},\hat Y_U \right] =
\left\{ \left(K^{\dagger}_ L\hat Y_DK_L \right)_{(nd)},\hat Y_U \right\} \eqno
(2.8a) $$
Notice that $ K^{\dagger}_ L\hat Y_DK_L $ is just $ Y_D $ in the basis where $
Y_U $ and $ {\rm SU}( {\rm 2})_{ {\rm weak}} $ are
diagonal (which will play a major r\^ole in the next sections) at each $ t. $
Analogously,
$$ \left[V^{\dagger}_ L { {\rm d} V_L \over {\rm d} t},\hat Y_D \right] =
\left\{ \left(K^{\dagger}_ L\hat Y_UK_L \right)_{(nd)},\hat Y_D \right\} \eqno
(2.8b) $$

Finally,
$$ { {\rm d} K_L \over {\rm d} t} = K_LU^{\dagger}_ L { {\rm d} U_L \over {\rm
d} t} - V^{\dagger}_ L { {\rm d} V_L \over {\rm d} t} K_L \eqno (2.9) $$

At this point we can take advantage of the strong hierarchy of the $ ( {\rm
mass})^2 $
eigenvalues for up-quarks on the one hand $ \left(m^2_t \gg  m^2_c \gg  m^2_u
\right), $ and for down-quarks on
the other hand
$ \left(m^2_b \gg  m^2_s \gg  m^2_d \right). $ Taking the components
of (2.8)
$$ \eqalign{ \left(U^{\dagger}_ L { {\rm d} U_L \over {\rm d} t} \right)_{ab}
& = \left({m^2_a + m^2_b \over m^2_a - m^2_b} \right) \left(K^{\dagger}_ L\hat
Y_DK_L \right)_{ab} \cr  & ( a,b=u,c,t;\ a\not= b) \cr \left(V^{\dagger}_ L {
{\rm d} V_L \over {\rm d} t} \right)_{ab} & = \left({m^2_a + m^2_b \over m^2_a
- m^2_b} \right) \left(K_L\hat Y_DK^{\dagger}_ L \right)_{ab} \cr  & (
a,b=d,s,b;\ a\not= b) \cr} \eqno (2.10) $$
one obtains the evolution of $ U_L $ and $ V_L $ from the $ ( {\rm mass})^2 $
hierarchy:
$$ \eqalign{ \left(U^{\dagger}_ L { {\rm d} U_L \over {\rm d} t} \right)_{ab}
& = \xi_{ ab} \left(K^{\dagger}_ L\hat Y_DK_L \right)_{ab} \cr
\left(V^{\dagger}_ L { {\rm d} V_L \over {\rm d} t} \right)_{ab} & = \xi_{ ab}
\left(K_L\hat Y_UK^{\dagger}_ L \right)_{ab} \cr} \eqno (2.11) $$
$$ \xi_{ ab}= \left\{ \matrix{ 1\ \ \ \ \ \ \ \ \ \ \left(m_a<m_b \right)
\hfill \cr 0\ \ \ \ \ \ \ \ \ \ (a=b) \hfill \cr -1\ \ \ \ \ \ \ \ \
\left(m_a>m_b \right) \hfill \cr} \right. \eqno  $$

Next order approximations in $ \left(m^2_s/m^2_b \right) $ and $
\left(m^2_c/m^2_t \right) $ are easily obtained (but
more difficult to integrate), though useless in physical applications. From
(2.9) and (2.11), we derive the RGE for the elements of the evolving KM
matrix. We fully use the hierarchy among masses of equal-charge quarks and
get the following results\footnote{$ ^3 $}{Notice that in spite of their real
aspect, Eqs.(2.12) and (2.13) are RGE for complex quantities.} \footnote{$ ^4
$}{For
simplicity the index $ L $ is omitted.}:
$$ \eqalign{{ {\rm d} \over {\rm d} t} {\rm ln} \ K_{bt} & = \zeta +
\left(\lambda^ 2_t+\lambda^ 2_b \right) \cr{ {\rm d} \over {\rm d} t} {\rm ln}
\ K_{bc} & = \zeta +\lambda^ 2_b \left\vert K_{bu} \right\vert^ 2 \cr{ {\rm d}
\over {\rm d} t} {\rm ln} \ K_{st} & = \zeta +\lambda^ 2_t \left\vert K_{dt}
\right\vert^ 2 \cr{ {\rm d} \over {\rm d} t} {\rm ln} \ K_{bu} & = \zeta
+\lambda^ 2_b \left\vert K_{bc} \right\vert^ 2 \cr{ {\rm d} \over {\rm d} t}
{\rm ln} \ K_{dt} & = \zeta +\lambda^ 2_t \left\vert K_{st} \right\vert^ 2 \cr
\zeta &  = - \left(\lambda^ 2_t + \lambda^ 2_b \right) \left\vert K_{bt}
\right\vert^ 2 \cr} \eqno (2.12) $$

Unitarity is easily checked.
The RGE for remaining elements of the KM matrix are more involved, so  we
take advantage of the strong hierarchy in the non-diagonal elements to write
them as follows:
$$ \eqalign{{ {\rm d} \over {\rm d} t} {\rm ln} \ K_{du} & \simeq  \lambda^
2_s \left\vert K_{su} \right\vert^ 2 + \lambda^ 2_b \left\vert K_{bu}
\right\vert^ 2 \cr  &  + \lambda^ 2_c \left\vert K_{dc} \right\vert^ 2 +
\lambda^ 2_t \left\vert K_{dt} \right\vert^ 2 \cr{ {\rm d} \over {\rm d} t}
{\rm ln} \ K_{sc} & \simeq  \lambda^ 2_s \left\vert K_{su} \right\vert^ 2 +
\lambda^ 2_b \left\vert K_{bc} \right\vert^ 2 \cr  &  + \lambda^ 2_c
\left\vert K_{dc} \right\vert^ 2 + \lambda^ 2_t \left\vert K_{st} \right\vert^
2 \cr{ {\rm d} \over {\rm d} t} {\rm ln} \ K_{dc} & \simeq  - \left(\lambda^
2_s+\lambda^ 2_c \right) \left\vert K_{sc} \right\vert^ 2 \cr  &  +
\left(\lambda^ 2_b+\lambda^ 2_t \right) \left\vert K_{dt} \right\vert^ 2 \cr{
{\rm d} \over {\rm d} t} {\rm ln} \ K_{su} & \simeq  - \left(\lambda^
2_s+\lambda^ 2_c \right) \left\vert K_{sc} \right\vert^ 2 \cr  &  +
\left(\lambda^ 2_b+\lambda^ 2_t \right) \left\vert K_{bu} \right\vert^ 2 \cr}
\eqno (2.13) $$

These results have been obtained by a different method in
Ref.$\lbrack$25$\rbrack$\footnote{$ ^5 $
}{And, previously, in Ref.$ [4] $ for $ \lambda^ 2_t\gg \lambda^ 2_b. $}. It
is
notice-worthy that the present approach clearly shows that the relevant
approximation is given by (2.11). The RG evolution of the KM matrix elements
in (2.13) is negligible  compared to those in (2.12). The latter are
easily integrated if one neglects the small matrix elements in the
r.h.s.\nobreak\ .
In practice, it is enough to put $ \left\vert K_{bt} \right\vert^ 2\simeq 1 $
to get
$$ \eqalign{{ K_{bu}(\Lambda) \over K_{bu} \left(\Lambda_ 0 \right)} & =
{K_{bc}(\Lambda) \over K_{bc} \left(\Lambda_ 0 \right)} = {K_{st}(\Lambda)
\over K_{st} \left(\Lambda_ 0 \right)} \cr  &  = {K_{dt}(\Lambda) \over K_{dt}
\left(\Lambda_ 0 \right)} = {\rm e}^{I_t+I_b} \cr I_t & = \int^ 0_t {\rm d} t\
\lambda^ 2_t(t)\ \ \ \ \ I_b = \int^ 0_t {\rm d} t\ \lambda^ 2_b(t) \cr} \eqno
(2.14) $$
while all other KM matrix elements have a negligible $ t $-dependence.
  Assume $ \lambda^ 2_t \gg  \lambda^
2_b, $ then, from the RGE
for $ \lambda^ 2_t $, one gets
$$ \eqalign{ {\rm e}^{-I_t} & = \left(1 - {\lambda^ 2_t(t) \over \lambda^ 2_{
{\rm crit}}(t)} \right)^{1/12} \cr{ 1 \over \lambda^ 2_{ {\rm crit}}(t)} & = 12
\int^ 0_t {\rm d} t\ \prod^{ }_ \alpha \left({g^2_\alpha( t) \over g^2_\alpha(
0)} \right)^{2C^u_\alpha /b_\alpha} \cr  & \left(b_3=3,\ b_2=-1,\ b_1=-11
\right) \cr} \eqno (2.15) $$
where $ \lambda^ 2_{ {\rm crit}}(t) $ is the $ \lambda^ 2_t $ value
corresponding to a Landau pole at $ t_0. $ Hence,
the RG evolution of the KM matrix can be dismissed in most problems, or
taken into account by simple approximations as in (2.14).

It goes without saying that everything can be repeated ipsis litteris for the
unitary transformations $ U_R, $ $ V_R, $ $ K_R $ on the right-handed quarks.

\vskip 17pt
\noindent {\bf 3. ANALYTIC SOFT TERMS}
\vskip 12pt

In this Section, we establish the RGE for the so-called soft analytic scalar
interactions. Those relevant for the phenomenology of flavour mixing are:
$$ H_1U^c\eta_ UQ + H_2D^c\eta_ DQ + {\rm h.c.} \ , \eqno (3.1) $$
where isospin indices are omitted, $ \eta_ U $ and $ \eta_ D $ are matrices in
the family
indices. The matrix elements of $ \eta_ U $ and $ \eta_ D $ are proportional
to the
supersymmetry breaking scale, hence to $ m_{3/2}. $ As discussed in the
introduction
(see also (3.5) below), it is simple to assume them to be also proportional to
the corresponding matrix elements of the Yukawa  couplings, $ \lambda_ U $ and
 $ \lambda_ D, $
respectively. The current assumption of \lq\lq universality\rq\rq , i.e.,
flavour
independence of the soft terms is usually stated by setting $ [15] $ $ \eta_
U=A\lambda_ U, $ $ \eta_ D=A\lambda_ D $
where $ A $ is $ 0 \left(m_{3/2} \right). $ Since flavour independence is
broken in the RGE by the
effect of Yukawa interactions, this is usually taken into account by
replacing $ A $ by a diagonal matrix and neglecting the scale dependence of
the
KM matrix. In more specific calculations (e.g., in that of the electric
dipole moment of the neutron in models where $ A $ is assumed to be universal)
the matrix character of the soft terms have to be more carefully manipulated
$ [27]. $
Nevertheless,  the soft couplings $ \eta_ U $ and $
\eta_ D $ are not
expected to be in general exactly flavour independent in supergravity theories.
In such  circumstances, one should choose some parametrisation of $ \eta_ U
$ and $ \eta_ D $
which simply embodies the assumed dependence on $ \lambda_ U $ and $ \lambda_ D
$ and still preserves
the generality of supergravity breaking terms. Obviously, it is always
possible to define a matrix $ A $ as $ \lambda^{ -1}\eta $ (or $ \eta
\lambda^{ -1}), $ but since these soft terms
and the Yukawa couplings link quarks with different $ q $-numbers $ (L $ and $
R $
states, respectively) let us scrutinise the RGE for $ \eta^ U, $
$$ \eqalign{{ {\rm d} \eta^ U \over {\rm d} t} & = \eta^ U \left[5\lambda^{
U\dagger} \lambda^ U+\lambda^{ D\dagger} \lambda^ D+3\ {\rm Tr}
\left(\lambda^{ U\dagger} \lambda^ U \right)-2C^u_\alpha g^2_\alpha \right]
\cr  &  + 2\lambda^ U \left[2\lambda^{ U\dagger} \eta^ U+3\ {\rm Tr}
\left(\lambda^{ U\dagger} \eta^ U \right)+\lambda^{ D\dagger} \eta^
D-2C^u_\alpha g^2_\alpha M_\alpha \right] \cr} \eqno (3.2) $$
and its analogous for $ \eta_ D, $ replacing $
U\rightleftharpoons D $ (and adding
 the leptonic couplings $ {\rm Tr} \ \lambda^{ L\dagger} \lambda^ L). $
Actually, the choice $ \eta^ U=\lambda^ UA^U $ and $ \eta^ D=\lambda^ DA^D $
is quite a satisfactory definition leading, e.g., to the following
equation for the  matrix  $ A^U $
$$ { {\rm d} A^U \over {\rm d} t}= 5A^UY_U+Y_UA^U+Y_DA^D+6\ {\rm Tr}
\left(A^UY_U \right)-4C^u_\alpha g^2_\alpha M_\alpha \eqno (3.3) $$
This choice has been used in the  literature\footnote{$ ^6 $}{See, e.g., Refs.
$ [4,8]. $}
in connection with
FCNC effects.

However, in the more general context of flavour dependent supergravity
couplings,  the choice in (3.3) is not the best in view of the structure of
the boundary conditions at the unification scale $ \Lambda_ 0, $ not far from
the Planck
mass. Indeed, consider an effective supergravity theory $ [15] $ defined by
some
Kahler potential $ K \left(X,...,H_1,H_2,U_i,Q_i,D_i... \right) $ $ (i=1,2,3 $
is the family index) and
a superpotential:
$$ W = H_1U^c\lambda^ U(X)Q + H_2D^c\lambda^ D(X)Q + ... \eqno (3.4) $$
where we have defined the chiral field $ X $ as the goldstino (i.e.,
supersymmetry breaking) direction, such that the auxiliary field $ F_X\sim 0
\left(m_{3/2} \right) $
(in units $ M_{ {\rm Planck}}=1). $ Then,
the trilinear coupling matrix $ \eta^ U $ will take the form $ [15,24]: $
$$ \eqalign{ \eta^ U_{ij} & = \left\{ \left({\partial \over \partial_ X} + {1
\over 2} {\partial K \over \partial X} \right)\lambda^ U_{ij}(X) + \Gamma^
H_{XH}\lambda^ U_{ij}(X) + \bar \Gamma^ k_{Xi}\lambda^ U_{kj}(X)+\Gamma^
k_{Xj}\lambda^ U_{ik}(X) \right\} F^X \cr \Gamma^ H_{XH} & = K^{HH^\ast}
{\partial \over \partial X} K_{H^\ast H} \cr\bar \Gamma^ k_{Xi} & = \bar
K^{k\ell^ \ast}  {\partial \over \partial X} \bar K_{\ell^ \ast i} \cr \Gamma^
k_{Xj} & = K^{k\ell^ \ast}  {\partial \over \partial X} K_{\ell^ \ast j} \cr}
\eqno (3.5) $$
where  $ K_{HH^\ast} , $ $ \bar K_{\ell^ \ast i} $ and $ K_{\ell^ \ast i} $
denote the effective Kahler metrics for $ H_1, $ $ U $ and
$ Q, $ respectively. This formula clearly shows how flavour mixing can be
introduced in the $ \eta $'s, through either the \lq\lq Kahler
connexions\rq\rq\ $ \Gamma ,\bar \Gamma $ or a
non-trivial dependence of the Yukawa  couplings on the goldstino scalar
partner, $ X. $

Therefore, we favour a different decomposition of the soft terms, as follows,
$$ \eqalign{ \eta^ U & =B^U_R\lambda_ U+\lambda_ UB^U_L \cr \eta^ D &
=B^D_R\lambda_ D+\lambda_ DB^D_L \cr} \eqno (3.6) $$
whose advantages become clear comparing with (3.5). This decomposition
is defined up to terms commuting with $ \lambda_ U $ $ (\lambda_ D, $
resp.) which can be
arbitrarily transferred from $ B_R $'s to $ B_L $'s, or vice-versa.
In our expressions there below, we shall equally share the family independent
parts between $ B_R $ and $ B_L. $
The RGE for the
$ B_R $'s and $ B_L $'s immediately follow from (3.2). However, it is
more convenient to treat these RGE in such a way as to easily utilise the
hierarchy of  Yukawa couplings, in particular, the large value of the top
mass.
The obvious idea is to keep $ \lambda_ U(\Lambda ) $ diagonal at any scale $
\Lambda $ by the
transformations $ U_R,U_L $ defined  in (2.1). This amounts to choosing a
rotating
basis because of (2.8a) and its analogue for $ U_R. $ Let us denote
the {\sl  diagonal\/} Yukawa
coupling matrix by $ \hat \lambda_ U(\Lambda )=U^{\dagger}_ R(\Lambda
)\lambda_ U(\Lambda )U_L(\Lambda ), $ and any matrix  in flavour space in
the corresponding basis by
$$ \hat M(\Lambda)  = U^{\dagger}( \Lambda) M(\Lambda) U(\Lambda) \eqno (3.7)
$$
where $ U=U_R $ or $ U_L, $ according as $ M $ acts on $ R $ or $ L $ states,
respectively. Hence,
one derives the RGE for $ \hat M(\Lambda ) $ in the rotating basis as
$$ { {\rm d}\hat M \over {\rm d} t} = U^{\dagger}( \Lambda)  { {\rm d} M \over
{\rm d} t} U(\Lambda) + \left[\hat M,\ U^{\dagger}  { {\rm d} U \over {\rm d}
t} \right] \eqno (3.8) $$
where the last term takes into account the rotating basis. From the
discussion of the previous Section, one knows that the effect of this last
term is quite small in the $ L $-sector, where the KM matrix in (2.11) is
known.
On the contrary, the corresponding KM matrix for the $ R $-sector is unknown
and
so are the angular velocities defined by the analogue of (2.11).

Let us now write down the RGE for the $ B^\prime {\rm s} $ in (3.6). Unless
otherwise stated,
all matrices are written in the rotating basis (diagonal $ \lambda_ U). $ (To
simplify
the notation we omit the \lq\lq $ \hat {\phantom{A}} $ \rq\rq\ symbol on these
matrices).
 From (3.6), (2.2) and
(3.2) one then derives the following RGE equations in the (real) diagonal $
\lambda_ U $
basis:
$$ \eqalign{{ {\rm d} B^U_R \over {\rm d} t} & = 2B^U_R\lambda^ 2_U+4\lambda^
2_UB^U_R+6\ {\rm Tr} \left(\lambda^ 2_UB^U_R \right)-2C^u_\alpha g^2_\alpha
M_\alpha \uniset  + \left[B^U_R,\ U^{\dagger}_ R { {\rm d} U_R \over {\rm d}
t} \right] \cr{ {\rm d} B^U_L \over {\rm d} t} & = 5B^U_L\lambda^ 2_U+\lambda^
2_UB^U_L+6\ {\rm Tr} \left(\lambda^ 2_UB^U_L \right)-2C^u_\alpha g^2_\alpha
M_\alpha \uniset  + \left[B^U_L,\ \lambda^{ \dagger}_ D\lambda_ D
\right]+2\lambda^{ \dagger}_ DB^D_R\lambda_ D \cr  &  + 2\lambda^{ \dagger}_
D\lambda_ DB^D_L + \left[B^U_L,U^{\dagger}_ L { {\rm d} U_L \over {\rm d} t}
\right] \cr} \eqno (3.9) $$
$$ \eqalign{{ {\rm d} B^D_R \over {\rm d} t} & = 2B^D_R\lambda^{ \dagger}_
D\lambda_ D+4\lambda^{ \dagger}_ D\lambda_ DB^D_R+ \left[6\ {\rm Tr}
\left(\lambda^{ \dagger}_ D\lambda_ DB^D_R \right)-2C^d_\alpha g^2_\alpha
M_\alpha \right]\uniset  + \left[B^D_R,\ U^{\dagger}_ R { {\rm d} U_R \over
{\rm d} t} \right] \cr{ {\rm d} B^D_L \over {\rm d} t} & = 5B^D_L\lambda^{
\dagger}_ U\lambda_ D+\lambda^{ \dagger}_ U\lambda_ DB^D_L+ \left[6\ {\rm Tr}
\left(\lambda^{ \dagger}_ D\lambda_ DB^D_L \right)2C^d_\alpha g^2_\alpha
M_\alpha \right]\uniset  + \left[B^D_L,\ \lambda^ 2_U \right]+2\lambda^
U_DB^U_R\lambda^ U \cr  &  + 2\lambda^ 2_UB^U_L + \left[B^D_L,U^{\dagger}_ L {
{\rm d} U_L \over {\rm d} t} \right] \cr} \eqno  $$
where $ M_\alpha $ are the gaugino masses and the terms proportional to the
identity
matrix $\uniset$ have been equitably distributed between $ B_L $ and $ B_R. $
The last term in
each equation corresponds to the action of an element of the $ {\rm SU(3)} $
algebra
whose expression is given by (2.11). Notice that $ U^{\dagger}_ L \left( {\rm
d} U_L/ {\rm d} t \right) $ is known in terms
of the KM matrix but $ U^{\dagger}_ R \left( {\rm d} U_R/ {\rm d} t \right) $
is unknown.

Let us now illustrate some of the feature of (3.9) by a simple, yet
possibly realistic, case where the RGE can be solved analytically. This is so
if all Yukawa couplings can be neglected but the top one, $ \lambda_ t. $ This
approximation which kills many interesting terms in the RGE above, allows
one to compare with previous work\footnote{$ ^7 $}{Analytic solutions
without
family mixing in the  $ \lambda_ t $-dominance approximation can be found in $
[4] $ and $ [28] $
with universal supergravity couplings.}
in the literature
where family mixing is neglected. Indeed, the solutions
of (3.9) in this approximation are as follows:
$$ \eqalign{ \left(B^U_L+B^U_R \right)_{ii}(t) & = \left(B^U_L+B^U_R
\right)_{ii}(0) - 4 \sum^{ }_ \alpha C^u_\alpha g^2_\alpha tM_0 \cr  &  -{\rho
\over 2} \left(1+\delta_{ i3} \right) \left[ \left(B^U_L+B^U_R
\right)_{33}(0)+\xi M_0 \right] \cr} $$
$$ \eqalign{ \left(B^U_R \right)_{ij}(t) & = {\rm e}^{-\alpha^ R_{ij}I_t}
\left(B^U_R \right)_{ij}(0) \cr \left(B^U_L \right)_{ij}(t) & = {\rm
e}^{-\alpha^ L_{ij}I_t} \left(B^U_L \right)_{ij}(0) \cr} $$
$$ \displaylines{ \alpha^ R_{23}=\alpha^ R_{13} = 2\ \ \ \ \alpha^ L_{23} =
\alpha^ L_{13} = 5 \cr \alpha^ R_{32}=\alpha^ R_{31} = 4\ \ \ \ \alpha^ L_{32}
= \alpha^ L_{31} = 1 \cr \alpha^ R_{12}=\alpha^ R_{21} = \alpha^ L_{12} =
\alpha^ L_{21} = 0 \cr} $$
$$ \eqalign{ \left(B^D_R \right)_{ij}(t) & = \left(B^D_R
\right)_{ij}(0)-2\delta_{ ij} \sum^{ }_ \alpha C^d_\alpha g^2_\alpha tM_0 \cr
&  \ \ \ \ \ \ (i,j = 1,2,3) \cr \left(B^D_L \right)_{ii}(t) & = \left(B^D_L
\right)_{ii}(0)-2 \sum^{ }_ \alpha C^d_\alpha g^2_\alpha tM_0 -{1 \over 6}
\rho \delta_{ i3} \left[ \left(B^U_R+B^U_L \right)_{33}(0)+\xi M_0 \right]
\cr} $$
$$ \eqalign{ \left(B^D_L \right)_{12}(t) & = \left(B^D_L \right)_{12}(0) \cr
\left(B^D_L \right)_{i3}(t) & = {\rm e}^{-I_t} \left(B^D_L \right)_{i3}(0)\ \
\ \ (i = 1,2) \cr \left(B^D_L \right)_{3i}(t) & = {\rm e}^{I_t} \left[
\left(B^D_L \right)_{3i}(0)-2I \left(B^U_L \right)_{3i}(0) \right]\ , \cr}
\eqno (3.10) $$
where $ I_t $ and $ {\rm e}^{-I_t} $ are defined in (2.15), $ M_0 $ is the
gaugino mass at the
unification scale $ (t=0), $ and $ \rho $ and $ \xi $ are defined by:
$$ \eqalign{ \rho &  = {\lambda^ 2_t(t) \over \lambda^ 2_{ {\rm crit}}(t)} = 1
- {\rm e}^{-12I_t} \cr \xi &  = -12\lambda^ 2_{ {\rm crit}}t-1\ . \cr} \eqno
(3.11) $$
Notice that $ {\rm e}^{I_t}\simeq 1, $ so that the sizeable effect in the $
B^D_L, $ $ B^D_R $ matrix elements
are the gauge contributions in the diagonal and the top trilinear term in $
\left(B^D_L \right)_{33}. $

The consistently complex character of all the relations in this section is
notice-worthy.

\vskip 17pt
\noindent {\bf 4. SCALAR MASS MATRICES}
\vskip 12pt
The RGE for the scalar $ ( {\rm mass})^2 $ soft terms are available in the
literature in
 full matrix form $ [4,8]. $ However the universality hypothesis has been
used in the solution $ [28]. $ The one-loop RGE for the relevant masses are,
$$ \eqalignno{{ {\rm d} m^2_Q \over {\rm d} t} & = \left \{ \lambda^{ \dagger}_
U\lambda_ U+\lambda^{ \dagger}_ D\lambda_ D,m^2_Q \right \} +2
\left(m^2_{H_1}\lambda^{ \dagger}_ U\lambda_ U+m^2_{H_2}\lambda^{ \dagger}_
D\lambda_ D+\lambda^{ \dagger}_ Um^2_U\lambda_ U+\lambda^{ \dagger}_
Dm^2_D\lambda_ D \right. &  \cr  & \left.+\eta^{ \dagger}_ U\eta_ U+\eta^{
\dagger}_ D\eta_ D \right)-8C^Q_\alpha g^2_\alpha M^2_\alpha  &  \cr{ {\rm d}
m^2_U \over {\rm d} t} & =2 \left\{ \lambda_ U\lambda^{ \dagger}_ U,
m^2_U \right\} +4 \left( m^2_{H_1}\lambda_ U\lambda^{ \dagger}_ U+\lambda_
Um^2_Q\lambda^{ \dagger}_ U+\eta_ U\eta^{ \dagger}_ U \right)-8C^U_\alpha
g^2_\alpha
M^2_\alpha +... & (4.1) \cr} $$
$$ \eqalignno{{ {\rm d} m^2_{H_1} \over {\rm d} t} & =6  {\rm Tr}
\left(m^2_{H_1}\lambda^{ \dagger}_ U\lambda_ U+m^2_Q\lambda^{ \dagger}_
U\lambda_ U+m^2_U\lambda_ U\lambda^{ \dagger}_ U+\eta^{ \dagger}_ U\eta_ U
\right)-8C^H_\alpha g^2_\alpha M^2_\alpha +... &  \cr} $$
where $ {\rm Tr} $ is the trace on flavour indices. The RGE for $ m^2_D $ and
$ m^2_{H_2} $ are obtained
from those of $ m^2_U $ and $ m^2_H, $ respectively by the exchanges $ U
\longleftrightarrow D, $ $ H_1 \longleftrightarrow H_2 $ and
inclusion of the leptonic terms. Because of  non-universality, there is the
additional term $ Y_iS $ in the RGE, where,
$$ S(t)={g^2_1(t) \over g^2_1(0)} {\rm Tr} \left(Ym^2 \right) \eqno (4.2) $$
The last term in each RGE takes into account the change of the basis at each
energy if, for instance, $ \lambda_ U $ is kept diagonal. Let us then display
the
solutions of the above RGE under the assumption that all Yukawa couplings but
$ \lambda^ 2_t $ are neglected, consistently with (3.10). In the basis in
which $ \lambda_ U $ is
diagonal, we obtain for those hermitian matrices:
$$ \eqalignno{ \left(m^2_Q \right)_{33}(t) & = \left(m^2_Q
\right)_{33}(0)+{g^2_1 \over 3}t\ S(0)-{\rho \over 6}K-\gamma^ Qt\ M^2_0 &
\cr \left(m^2_U \right)_{33}(t) & = \left(m^2_U \right)_{33}(0)-{4 \over
3}g^2_1t\ S(0)-{\rho \over 3}K-\gamma^ Ut\ M^2_0 &  \cr
 m^2_D(t) & =m^2_D(0)+ \left({2g^2_1 \over 3}t\ S(0)-\gamma^ Dt\
M^2_0 \right)\uniset &  \cr \left(m^2_Q \right)_{ij}(t) & = \left(m^2_Q
\right)_{ij}(0)+{g^2_1 \over 3}t\ S(0)\delta_{ ij} + \left( {\rm e}^{-2I_t}-1
\right) \left(B^{U\ast}_ L \right)_{3i} \left(B^U_L \right)_{3j}(0)-\gamma^
Qt\ M^2_0\delta_{ ij} &  \cr \left(m^2_U \right)_{ij}(t) & = \left(m^2_U
\right)_{ij}(0)-{4 \over 3}g^2_1t\ S(0)\delta_{ ij} + \left( {\rm e}^{-4I_t}-1
\right) \left(B^U_R \right)_{i3} \left(B^{U\ast}_ R \right)_{j3}(0)-\gamma^
Ut\ M^2_0\delta_{ ij} &  \cr \left(m^2_Q \right)_{i3}(t) & = {\rm e}^{-I_t}
\left(m^2_Q \right)_{i3}(0)-{\rho \over 6}A_0 \left(B^{U\ast}_ L
\right)_{3i}(0)
&  \cr \left(m^2_U \right)_{i3}(t) & = {\rm e}^{-2I_t} \left(m^2_U
\right)_{i3}(0)-{\rho \over 3}A^\ast _0 \left(B^U_R \right)_{i3}(0) & (4.3)
 \cr} $$
where $ i,j=1,2. $ We have defined the following quantities:
$$ \eqalignno{ A_0 & = \left(B^U_R+B^U_L \right)_{33}(0)+\xi \ M_0 &  \cr
{\phantom {A}} K & =
\left(m^2_Q+m^2_U+m^2_{H_2}\uniset \right)_{33}(0)+(1-\rho ) \vert A_0 \vert
^2  -\xi ^2M^2_0 & \cr & -8 \sum^{ }_ \alpha C^u_\alpha g^2_\alpha( \xi +1)t\
M^2_0 &  \cr
\gamma^ Q & ={32 \over 3}g^2_3 \left(1+3g^2_3t \right)+6g^2_2 \left(1-g^2_2t
\right)+{2 \over 9}g^2_1 \left(1-11g^2_1t \right) &  \cr \gamma^ U & ={32
\over 3}g^2_3 \left(1+3g^2_3t \right)+{32 \over 9}g^2_1 \left(1-11g^2_1t
\right) &  \cr \gamma^ D & ={32 \over 3}g^2_3 \left(1+3g^2_3t \right)+{8 \over
9}g^2_1 \left(1-11g^2_1t \right) & (4.4) \cr} $$

As already discussed in the literature $ [3,4,8] $, starting from universal
soft
terms, the flavour dependence due to the RG running  only appears in the
diagonal matrix elements. In the general case instead, the flavour
dependence of the $ ( {\rm mass})^2 $ terms gets contributions from the $
B^U_L, $ $ B^U_R $ matrix
elements. Also in this Section, the complex nature of the various parameters
is explicitly respected.

Let us estimate the impact of the running of the soft terms from the
unification scale down to the $ {\rm SU(2)\times U(1)} $ symmetry breaking
scale. Assuming $ h^2_t\simeq 1 $
at $ \Lambda \sim 0 \left(M_Z \right), $ one gets the numerical values:
$$ \eqalignno{ t=-.2 & \ \ \ \ \rho =.8\ \ \ \ \ \ \xi =2 &  \cr {\rm
e}^{-I_t}=.87 & \ \ \ \ I_t=.13 &  \cr} $$
$$ \eqalignno{ 2 \sum^{ }_{ } C^u_\alpha g^2_\alpha t & =2 \sum^{ }_{ }
C^d_\alpha g^2_\alpha t=-1.9 &  \cr \gamma^ Qt=6.7 & \ \ \ \ \ \ \gamma^
Ut=\gamma^ Dt=6.3 & (4.5) \cr} $$
The substitution of these numbers in (3.11), (4.3) and (4.4) gives the
following relations,
$$ \eqalignno{ \left(B^U_L+B^U_R \right)_{ii}(t) & = \left(B^U_L+B^U_R
\right)_{ii}(0)+4M_0-.4 \left(1+\delta_{ i3} \right)A_0 &  \cr
\left(B^D_R+B^D_L \right)_{ii}(t) & = \left(B^D_L+B^D_R
\right)_{ii}(0)+4M_0-.13\delta_{ i3}A_0\ \ \ \ \ \ (i=1,2,3) &  \cr
\left(B^U_L \right)_{i3}(t) & =.5 \left(B^U_L \right)_{i3}(0)\ \ \ \ \ \ \ \
(i=1,2) &  \cr \left(m^2_Q \right)_{33}(t) & = \left(m^2_Q
\right)_{33}(0)-.4\bar m^2_0-.025A^2_0+5.7M^2_0 &  \cr \left(m^2_Q
\right)_{ij}(t) & = \left(m^2_Q \right)_{ij}(0)+6.7M^2_0\delta_{ ij}-.25
\left(B^{U\ast}_ L \right)_{3i} \left(B^U_L \right)_{3j}(0) &  \cr \left(m^2_Q
\right)_{i3}(t) & =.87 \left(m^2_Q \right)_{i3}(0)-.13A_0 \left(B^{U\ast}_ L
\right)_{3i}(0)\ \ \ \ \ \ (i,j=1,2) &  \cr} $$
$$ \eqalignno{ \left(m^2_U \right)_{33}(t) & = \left(m^2_U
\right)_{33}(0)-.8\bar m^2_0-.05A^2_0+4.3M^2_0 & (4.6) \cr \left(m^2_U
\right)_{ij}(t) & = \left(m^2_U \right)_{ij}(0)+6.3M^2_0\delta_{ ij}-.4
\left(B^{U\ast}_ R \right)_{i3} \left(B^U_R \right)_{j3} &  \cr \left(m^2_U
\right)_{i3}(t) & =.87 \left(m^2_Q \right)_{i3}(0)-.13A^\ast_ 0 \left(B^U_R
\right)_{i3}(0) &  \cr} $$
where,
$$ \eqalignno{ m^2_0 & ={1 \over 3} \left(m^2_Q+m^2_U+m^2_{H1} \right)_{33}(0)
&  \cr A_0 & = \left(B^U_L+B^U_R \right)_{33}(0)+2M_0 &  \cr} $$
and $ M_0 $ is the mass of the gauginos at the unification scale,
assumed to be all equal.

Since FCNC effects from the non-universality of $ m^2_U, $ $ m^2_D, $ $ m^2_Q
$ are expected to be
relatively small, it is convenient to expand the contributions to the
possible FCNC processes in powers of the following matrices:
$$ \eqalignno{ \left(\delta^ U_R \right)_{ij} & ={ \left(m^2_U \right)_{ij}
\over\bar m^2_U}-\delta_{ ij} &  \cr \left(\delta^ U_L \right)_{ij} & ={
\left(m^2_Q \right)_{ij} \over\bar m^2_Q}-\delta_{ ij} & (4.7) \cr} $$
where
$$ \eqalignno{\bar m^2_Q & ={1 \over 3} {\rm Tr\ } m^2_Q(0)+\gamma^ Q\vert
t\vert M^2_0+{g^2_1 \over 3}t\ S(0) &  \cr\bar m^2_U & ={1 \over 3} {\rm Tr\ }
m^2_U(0)+\gamma^ U\vert t\vert M^2_0-{4 \over 3}g^2_1t\ S(0) &  \cr} $$
(i.e., only the gauge contribution in RGE are included in the definition of
these \lq\lq average\rq\rq\ masses). By definition, these matrices are defined
in the
diagonal $ \lambda_ U $ basis. Instead, for the corresponding $ \delta^ D_R $
and $ \delta^ D_L, $ it is more
convenient to go to the diagonal $ \lambda_ D $ basis through KM rotations, $
K_L, $ $ K_R. $
Hence, we define the following matrices:
$$ \eqalignno{ \left(\tilde \delta^ D_R \right)_{ij} & ={
\left(K_Rm^2_DK^{\dagger}_ R \right)_{ij} \over\bar m^2_D}-\delta_{ ij} &  \cr
\left(\tilde \delta^ D_L \right)_{ij} & ={ \left(K_Lm^2_QK^{\dagger}_ L
\right)_{ij} \over\bar m^2_Q}-\delta_{ ij} & (4.8) \cr} $$
where the \lq\lq$ \sim $\rq\rq\ symbol indicates the diagonal $ \lambda_ D $
basis for the matrices.
Unfortunately, $ \tilde \delta^ D_R $ depends on the unknown $ K_R $ matrix.
The matrices defined in
(4.6) are those usually compared to experimental data on the various FCNC
processes. Thus, there are several upper bounds in the literature obtained by
considering a variety of physical transitions $ [8,10,11]. $ However these
upper
bounds correspond to different assumptions and approximations, and also
depend on the overall squark and gaugino masses.
\vskip 17pt
\noindent {\bf 5. AN EXAMPLE: THE NEUTRON E.D.M.}
\vskip 12pt

The neutron electric dipole moment is a precious parameter in fundamental
physics  as it is strongly suppressed in the Standard Model,
 on the contrary, it is sensitive to new
sources of CP violations. Among the possible physical phases in
supersymmetric theories, some are more relevant for the discussion of the
neutron e.d.m., namely, the relative phases between the matrix elements in
the soft analytic scalar couplings and the gaugino masses $ [26,27]. $ We
choose the
usual conventions and redefine the fields so that gaugino masses are real. In
this Section, we choose to work in the basis where the down-quark mass matrix
is diagonal and real. With respect to the definition in the previous
Sections, we have to perform KM rotations to define, e.g.,
$$ \eqalignno{\tilde m^2_Q & =K_Lm^2_QK^{\dagger}_ L &  \cr\tilde m^2_D &
=K_Rm^2_DK^{\dagger}_ R &  \cr\tilde \eta_ D & =K_R\eta_ DK^{\dagger}_ L &
(5.1) \cr} $$

Since the calculation of the gluino-squark contribution to the e.d.m. $ D_n $
of
the neutron has been already presented many times in the literature, we just
quote the simple final expression for the e.d.m. of the $ d $-quark:
$$ \eqalignno{ D_d & =-{8 \over 9} {\rm e}{\alpha_ 3 \over 4\pi} M^3_3V_2 {\rm
\ Im\ } \Gamma_{ dd} &  \cr \Gamma_{ dd} & = \int^{ }_{ }{ k^2 {\rm d} k^2
\over \left[k^2+M^2_3 \right]} \left({1 \over k^2+\tilde m^2_{D_R}}\tilde
\eta_ D{1 \over k^2+\tilde m^2_{D_L}} \right)_{dd} & (5.2) \cr} $$
where $ M_3 $ is the gluino mass, only the lowest order in the analytic
trilinear
coupling is retained, the phase in the \lq\lq$ \mu -term $\rq\rq\ is
neglected for simplicity, and the matrices $ \tilde m^2_R $
and $ \tilde m^2_L $ are defined as:
$$ \eqalignno{\tilde m^2_{D_R} & =\tilde m^2_D+ \left(\tilde \lambda_ DV_2
\right)^2-{M^2_Z \over 3}\ {\rm sin}^2\ \theta_ W\ {\rm cos}^2\ \beta &
\cr\tilde m^2_{D_L} & =\tilde m^2_Q+ \left(\tilde \lambda_ DV_2
\right)^2-M^2_Z \left({1 \over 2}-{ {\rm sin}^2\ \theta_ W \over 3} \right)\
{\rm cos}^2\ \beta &  (5.3) \cr} $$
$$ {\rm tan} \ \beta ={V_1 \over V_2}\ ,\ \ \ \ V_i= \left\langle H_i
\right\rangle \ . \eqno  $$
(All matrices are defined in the diagonal $ \lambda_ D $ basis).

In order to extract the dominant contributions in the presence of flavour
dependent supergravity couplings, let us consider a simple approximation. We
consider a Taylor expansion of the propagators $ \left(k^2-\tilde m^2_{D_R}
\right)^{-1} $ and $ \left(k^2-\tilde m^2_{D_L} \right)^{-1} $
around some mean value $ \bar m^2, $ an average over the squark and gluino
masses, and
then we integrate on $ k^2. $ By keeping only the dominant term of each kind,
using the notation introduced in (4.6), one ends up with the following
expression:
$$ \eqalignno{ {\rm Im} \ \Gamma_{ dd} & \simeq{ 1 \over\bar m^6} \left\{{
\lambda_ d \over 12}\ {\rm Im} \left(\tilde B^D_R+\tilde B^D_L
\right)_{dd}-{\lambda_ i \over 20}\ {\rm Im} \left[ \left(\tilde \delta^ D_R
\right)_{di} \left(\tilde B^D_L \right)_{id}+ \left(\tilde B^D_R \right)_{di}
\left(\tilde \delta^ D_L \right)_{id} \right] \right. &  \cr  &
\left.-{\lambda_ i \over 30}\ {\rm Im} \left[ \left(\tilde \delta^ D_R
\right)_{di} \left(\tilde \delta^ D_L \right)_{jd} \left(\tilde B^D_L
\right)_{ij}+ \left(\tilde \delta^ D_R \right)_{dj} \left(\tilde \delta^ D_L
\right)_{id} \left(\tilde B^D_R \right)_{ji} \right] \right\} &  (5.4) \cr} $$
where $ i=b,s; $ $ j=b,s,d. $ Because of the Yukawa coupling
hierarchy, all  terms
in (5.4) can be equally relevant. Indeed, let us consider the upper bounds
obtained in Refs.$ [8,10,11] $ for the squark mass differences, assuming $
\bar m\sim 0(1\ {\rm TeV}). $
They are as follows:
$$ \eqalignno{ \left(\tilde \delta^ D_L \right)_{sd} <.1\ \ \ \ \  & \ \ \ \ \
\
\left(\tilde \delta^ D_R \right)_{ds}<.1 \cr \left(\tilde \delta^ D_L
\right)_{bd} <.3 \ \ \ \ \ & \ \ \ \ \ \ \left(\tilde \delta^ D_R
\right)_{db}<.3 \cr \left(\tilde \delta^ D_R \right)_{ds} \left(
\tilde \delta^ D_L
\right)_{sd} & <4\times 10^{-5} \cr \left(\tilde \delta^ D_R \right)_{db}
\left(\tilde \delta^ D_L \right)_{bd} & <5\times 10^{-3} & (5.5) \cr} $$
Combining these values with the quark mass ratios, we get the following
estimates for the relative magnitude of the various terms appearing in
(5.4)\footnote{$ ^8 $}{
The estimates in brackets are obtained from the bounds on the products $
\delta^ D_L\delta^ D_R $
and the assumption $ \delta^ D_L\sim \delta^ D_R. $}:
$$ \eqalignno{{ 12\ m_s \over 20\ m_d} \left(\tilde \delta^ D_{L(R)}
\right)_{ds} & \sim 0(1)\ (0(\cdot 07)) &  \cr{ 12\ m_b \over 20\ m_d}
\left(\tilde \delta^ D_{L(R)} \right)_{db} & \sim 0(70)\ (0(20)) &  \cr{ 12\
m_s \over 30\ m_d} \left(\tilde \delta^ D_R \right)_{ds} \left(\tilde \delta^
D_L \right)_{sd} & \sim 0 \left(3\times 10^{-4} \right) &  \cr{ 12\ m_b
\over 30\ m_d} \left(\tilde \delta^ D_R \right)_{db} \left(\tilde \delta^ D_L
\right)_{bd} & \sim 0(1) & (5.6) \cr} $$

Though the coefficients of $ \left(\tilde B^D_L \right)_{bd}, $ $ \left(\tilde
B^D_R \right)_{db} $ are allowed to be rather large,
upper bounds are also available for these matrix elements. A rough upper
bound is obtained by assuming $ \left(B^D_{L(R)} \right)_{dd}\sim 0 \left(\bar
m \right), $ where $ \bar m $ is the average mass in
(5.4) and then imposing  bounds on $ (\eta)_{ bd}, $ $ (\eta)_{ db} $
obtained from studies
of supersymmetric
effects in $ K $ and $ B $ physics $ [8,10,11]. $
However, once translated in terms of the parameters in (5.4), these bounds
just require $ \left\vert \left(\tilde B^D_{L(R)} \right)_{db} \right\vert
\sim 0 \left( \left\vert \left(\tilde B^D_{L,R} \right)_{dd} \right\vert
\right). $ Thus, terms proportional to $ m_b $ can
even dominate  the first, more familiar, term.

We have dealt with $ D_d, $ the e.d.m. of the $ d $-quark
as some comparison of the different contributions can be performed.
 However, the
importance of terms proportional to heavy quark Yukawa couplings is even more
obvious if one calculates $ D_u, $ the e.d.m. of the $ u $-quark. In this
case we
work directly in the diagonal $ \lambda_ U $ basis used in the previous
Sections. The
analogue of (5.2) reads
$$ \eqalignno{ D_u & =+{16 \over 9} {\rm e}{\alpha_ 3 \over 4\pi} M^3_3V_1\
{\rm Im} \ \Gamma_{ uu} &  \cr \Gamma_{ uu} & = \int^{ }_{ }{ k^2 {\rm d} k^2
\over \left[k^2+M^2_3 \right]^3} \left({1 \over k^2+m^2_{U_R}}\eta_ U{1 \over
k^2+m^2_{U_L}} \right)_{uu} &  \cr m^2_{U_R} & =m^2_U+ \left(\lambda_ UV_1
\right)^2+{2M^2_Z \over 3}\ {\rm sin}^2\ \theta_ W\ {\rm cos} \ 2\beta &  \cr
m^2_{U_L} & =m^2_Q+ \left(\lambda_ UV_1 \right)^2+M^2_Z \left({1 \over 2}-{2\
{\rm sin}^2\ \theta_ W \over 3} \right)\ {\rm cos} \ 2\beta &  (5.7) \cr} $$

Then the same approximations as in (5.4) (but, in this case, they could be
poorer) yield,
$$ \eqalignno{ V_1\ {\rm Im} \ \Gamma_{ uu} & \simeq{ 1 \over\bar m^6}
\left\{{ m_u \over 12}\ {\rm Im} \left(B^U_R+B^U_L \right)_{uu}-{m_i \over
20}\ {\rm Im} \left[ \left(\delta^ U_R \right)_{ui} \left(B^U_L \right)_{iu}+
\left(B^U_R \right)_{ui} \left(\delta^ U_L \right)_{iu} \right] \right. &  \cr
 & \left.-{m_i \over 30}\ {\rm Im} \left[ \left(\delta^ U_R \right)_{ui}
\left(\delta^ U_L \right)_{ju} \left(B^U_L \right)_{ij}+ \left(\delta^ U_R
\right)_{uj} \left(\delta^ U_L \right)_{iu} \left(B^U_R \right)_{ji} \right]
\right\} &  (5.8) \cr} $$

Estimates like those in (5.6) are only possible for $ (uc) $ or $ (cu) $
matrix
elements. Using results from $ [8,10,11] $ we obtain the very poor constraints,
$$ \eqalignno{{ 12\ m_c \over 20m_u} \left(\delta^ U_{L(R)} \right)_{uc} &
\sim 0(14)\ (0(6)) &  \cr{ 12m_c \over 30m_u} \left(\delta^ U_R \right)_{uc}
\left(\delta^ U_L \right)_{cu} & \sim 0(.2) & (5.9)  \cr} $$

The largest contributions are certainly those proportional to $ m_t, $
the
top mass. Let us scrutinise the RGE of the relevant soft parameters in the
approximation of (3.10). First, we notice that $ \left(B^U_R \right)_{ut} $
and $ \left(B^U_L \right)_{tu} $ are only
slightly renormalised even for $ \lambda_ t $ near the critical (\lq\lq
fixed-point\rq\rq ) value
(while the transposed matrix elements get more reduced). Instead the $
\left(B^U_L+B^U_R \right)(33) $
diagonal element can be strongly reduced in the large $ \lambda_ t $ limit so
that
non-diagonal $ \left(B^U_L \right) $ and $ \left(B^U_R \right) $ can dominate.
This is even enhanced by the large
contributions to $ \left(\delta^ U_R \right), $ $ \left(\delta^ U_L \right) $
diagonal terms for large $ \lambda^ 2_t. $

Therefore, the contributions to $ D_d $ and, especially, $ D_u $ from terms
proportional
to $ m_b $ and $ m_t, $ respectively, could be larger than those obtained by
neglecting
the squark mass splitting in the propagator, which are proportional to light
quark Yukawas. However, the whole discussion remains merely academic in the
absence of a specific model for  supergravity parameters, including the
phases in the soft terms. The minimalist assumption is to require universal
supergravity boundary conditions and the KM phase as the only source of CP
violation. This has been shown to yield extremely small contributions to the
quark e.d.m. $ [27]. $ Another rough estimate $ [26] $ consists in setting $
\left(B_R+B_L \right)\sim M_3\sim\bar m $
to obtain for the phase $ \phi $ of $ \left(B_R+B_L \right), $ $ {\rm sin} \
\phi < \left(700\ {\rm MeV} /\bar m \right)^2, $ from the
experimental limit on $ D_N. $ These two extreme cases clearly show that the
effects could be important compared to the Standard Model prediction
where $ D_N $ is very strongly suppressed.

\vskip 17pt
\noindent {\bf 6. MAXIMUM FLAVOUR MIXING IN SUPERGRAVITY}
\vskip 12pt

In this Section we aim at studying a model with maximal mixing at the
supergravity scale, in order
 to get some quantitative insight about the effects at low
energies. A natural model for flavour mixing would constrain both the Kahler
potential and the superpotential (Yukawa couplings) in terms of symmetries.
We do not know a realistic model of this sort and so we forget the naturalness
requirement. Thus, we  discuss a model designed to stress FCNC
 effects where the flavour dependence is introduced by hand
in the (field dependent) supergravity metrics.
For simplicity, the Yukawa couplings are assumed
to be left-right symmetric. The $ U $ and $ V $ matrices at the unification
scale $ \Lambda_{ {\rm GUT}} $
are defined from the Yukawa couplings at $ \Lambda_{ {\rm GUT}} $ by
(2.1) and the KM matrix is given by the relation
$$ \eqalignno{  V & =U\ K^{\dagger}. & (6.1) \cr} $$

Let us consider the real matrix
$$ U={1 \over \sqrt{ 6}} \left( \matrix{ \sqrt{ 3}  &    & 1  &    & \sqrt{ 2}
\cr   &    &    &    &   \cr - \sqrt{ 3}  &    & 1  &    & \sqrt{ 2} \cr   &
 &    &    &   \cr 0  &    & -2  &    & \sqrt{ 2} \cr} \right) \eqno (6.2) $$
(we neglect all phases in this Section). From (6.1) one gets the
following model for $ \lambda_ U \left(\Lambda_{ {\rm GUT}} \right): $
$$ \eqalignno{ \lambda_ U & ={\lambda_ t \over 3} \left( \matrix{
1+\varepsilon_ u+\varepsilon_ c  &    & 1-\varepsilon_ u+\varepsilon_ c  &
& 1-2\varepsilon_ c \cr   &    &    &    &   \cr 1-\varepsilon_ u+\varepsilon_
c  &    & 1+\varepsilon_ u+\varepsilon_ c  &    & 1-2\varepsilon_ c \cr   &
&    &    &   \cr 1-2\varepsilon_ c  &    & 1-2\varepsilon_ c  &    &
1+4\varepsilon_ c \cr} \right) &  \cr \varepsilon_ u & ={3\lambda_ u \over 2\
\lambda_ t}\ \ \ \ \ \ \ \ \ \ \varepsilon_ c={\lambda_ c \over 2\lambda_ t}
& (6.3) \cr} $$
This matrix is (almost) \lq\lq democratic\rq\rq , in the sense that all matrix
elements
are almost equal. In this sense the flavour mixing is  maximal. Then for the
sake of our qualitative discussion , we define $ V $ as in (6.1) with
$ K $ phenomenologically given.

The supergravity basis is defined by the diagonalization of the Kahler
matrix. In this basis, an honest assumption is to set all soft terms diagonal
at $ \Lambda_{ {\rm GUT}} $ (however, this is not necessarily so, as
discussed in Ref.[6]).  Thus, let us define
$$ \eqalignno{ \left(B^U_R \right)_{ab} & = \left({A^U \over 2}+b^U_{Ra}
\right)\delta_{ ab} &  \cr \left(B^U_L \right)_{ij} & = \left({A^U \over
2}+b^U_{Li} \right)\delta_{ ij} &  \cr \left(m^2_Q \right)_{ij} & = \left(\bar
m^2_Q+\mu^ 2_{Qi} \right)\delta_{ ij} &  \cr \left(m^2_U \right)_{ab} & =
\left(\bar m^2_U+\mu^ 2_{Ua} \right)\delta_{ ab} &  \cr \left(m^2_D
\right)_{ab} & = \left(\bar m^2_D+\mu^ 2_{Da} \right)\delta_{ ab} &  \cr  &
(i,j,a,b=1,2,3) & (6.4) \cr} $$

By performing the $ U $ rotation, the soft terms are transformed into the more
convenient diagonal $ \lambda_ U $ basis. In this basis $ B^U_R $ and $ B^U_L
$ take the form:
$$ \eqalignno{ B^U_R & = \left({A_U \over 2}+{ \sum^{ }_ ab^U_{Ra} \over 3}
\right)\uniset +{1 \over \sqrt{ 2}} \left( \matrix{ \sqrt{ 3}\Delta b^U_R  &
 & \Delta '  b^U_R  &    & \sqrt{ 2}\Delta ' b^U_R \cr   &    &    &    &   \cr
\Delta ' b^U_R  &    & - \sqrt{ 3}\Delta b^U_R  &    & \Delta b^U_R \cr   &
&    &    &   \cr \sqrt{ 2}\Delta ' b^U_R  &    & \Delta b^U_R  &    & 0 \cr}
\right) &  \cr \Delta b^U_R & =b^U_{R1}+b^U_{R2}-2b^U_{R3} &  \cr \Delta
'
b^U_R & =b^U_{R1}-b^U_{R2} & (6.5) \cr} $$
and the analogous expression for $ B^U_L $ with $ R \longrightarrow L. $
The corresponding relations for $ B^D_R, $ $ B^D_L $ are analogous with $ U
\longrightarrow D $ in the basis
where $ \lambda_ D $ is diagonal and thus they are transformed by a KM
rotation into the
diagonal $ \lambda_ U $ basis.

Similar expressions also result for the $ m^2 $ parameters, which can be
written
as:
$$ \eqalignno{ m^2_I & = \left(\bar m^2_I+{1 \over 3} \sum^{ }_ i\mu^ 2_{Ii}
\right)+{1 \over \sqrt{ 2}} \left( \matrix{ \sqrt{ 3}\Delta \mu^2_I  &    &
\Delta ' \mu^2_I  &    & \sqrt{ 2}\Delta '\mu^2_I \cr   &    &    &    &   \cr
\Delta '\mu^2_I  &    & - \sqrt{ 3}\Delta\mu^2_I  &    & \Delta\mu^2_I \cr   &
   &    &    &   \cr \sqrt{ 2}\Delta '\mu^2_I  &    & \Delta\mu^2_I  &    & 0
\cr} \right) &  \cr  &  (I=Q,U,D) & (6.6) \cr} $$

In order to proceed and fix some parameters in view of a quantitative
analysis, we consider a simple Ansatz for the quark quadratic terms in the
Kahler potential
$$ K \left(T,\Phi^ A,... \right)= \sum^{ }_ A{\Phi^ A\Phi^{ A\ast} \over( {\rm
Re} \ T)^{-n_A}}+... \eqno (6.7) $$
where $ \Phi^ A=Q^i,U^a,D^a. $ The moduli fields introduced in the
effective supergravity by the compactification of six dimensions in
string theory, are represented here by the overall modulus, denoted $T$.
The exponents $n_A$ are the modular weights that specify the behaviour
of each superfield $\Phi_A$ under modular transformations of the
compact manifold. We shall assume them to be negative integer numbers
$[17,18].$ Then, from the general supergravity
expressions:
$$ m^2_{AA\ast}= \left(F^MF^{N\ast} \right) \left(Z_{MN\ast} Z_{AA\ast}
-R_{MN\ast
AA\ast} \right) \eqno (6.8)  $$
$$ \eqalignno{ \eta_{ ABC} & =F^MD_M\lambda_{ ABC} &  \cr D_M & =Z^{-1}
\left(\partial_ M+{1 \over 2}\partial_ MK \right)Z &  \cr Z_{AB\ast} &
=\partial_ A\partial_{ B\ast} K & (6.9)  \cr} $$
one gets the soft scalar couplings and masses in
terms of the  supersymmetry breaking auxiliary fields $ F^M. $ Here we
consider
the simplest case   in which these auxiliary fields are
just $ F^S $ and $ F^T, $
corresponding to the  dilaton sector and to the overall modulus sector in the
effective supergravity theory obtained from the orbifold compactification
of superstrings. Accordingly, we introduce the Goldstino angle $ \theta $
$ [12] $, such that $ \tan \theta = F^S / F^T $, and neglect possible phases.
Then the gaugino masses at $ \Lambda_{ {\rm GUT}} $ are
all equal and given
by $ M^2_0= \left\vert F^S \right\vert^ 2, $ while for the other parameters
one obtains $ [12] $
$$ \eqalignno{ m^2_A & ={M^2_0 \over 3} \left(1+(1+n_A) {\rm \ cot}^2\ \theta
\right) &  \cr \left(B^U_R \right)_{ab} & =-M_0({1\over 2}+
{1\over \sqrt 3}({3\over 2}+{n_{H_1}\over 2}+n^{U}_a)\cot\theta) \delta_{ab} &
(6.10)
\cr} $$
with analogous expressions for  $ B^U_L, $ $ B^D_R, $ $
B^D_L. $

 These expressions can be
substituted in (4.6) to allow a numerical investigation. The integers
$n_i$ (modular weights) are free parameters. We choose to fix their
values $n_1=-1,\ n_2=-2,\ n_3=-3$ for the first, second and third
families. Modular weights of Higgs fields are chosen to be -2. These
choices are compatible with known results from superstring
compactification where $n_i$ is -2 for non oscillated twisted fields
of orbifold models. Anyway, this particular choice is only meant to
shift the degeneracy of $B'$s and $ m^2 $ 's at the unification scale
in a way which is qualitatively consistent with our expectations from
supergravity theory.  We impose that masses of squarks
at the compactification scale are all positive. This
entails that:
$$\tan ^2  \theta > -(n_m+1)\eqno (6.11)$$
where $n_m$ is the lowest modular weight. In our case $\tan ^2 \theta
>  2 $. As a matter of fact,
 this is not strictly necessary since a negative squark $ {\rm mass}^2 $ at the
unification scale will be turned positive (mostly) by strong interaction
radiative corrections, cf. $ (4.6). $ But if one assumes similar modular
weights for the lepton sector, then $ (6.11) $ ensures that sneutrino and
selectron masses are consistent with experimental bounds.

\vskip -3.0 cm
\epsfysize=12.cm{\centerline{\epsfbox{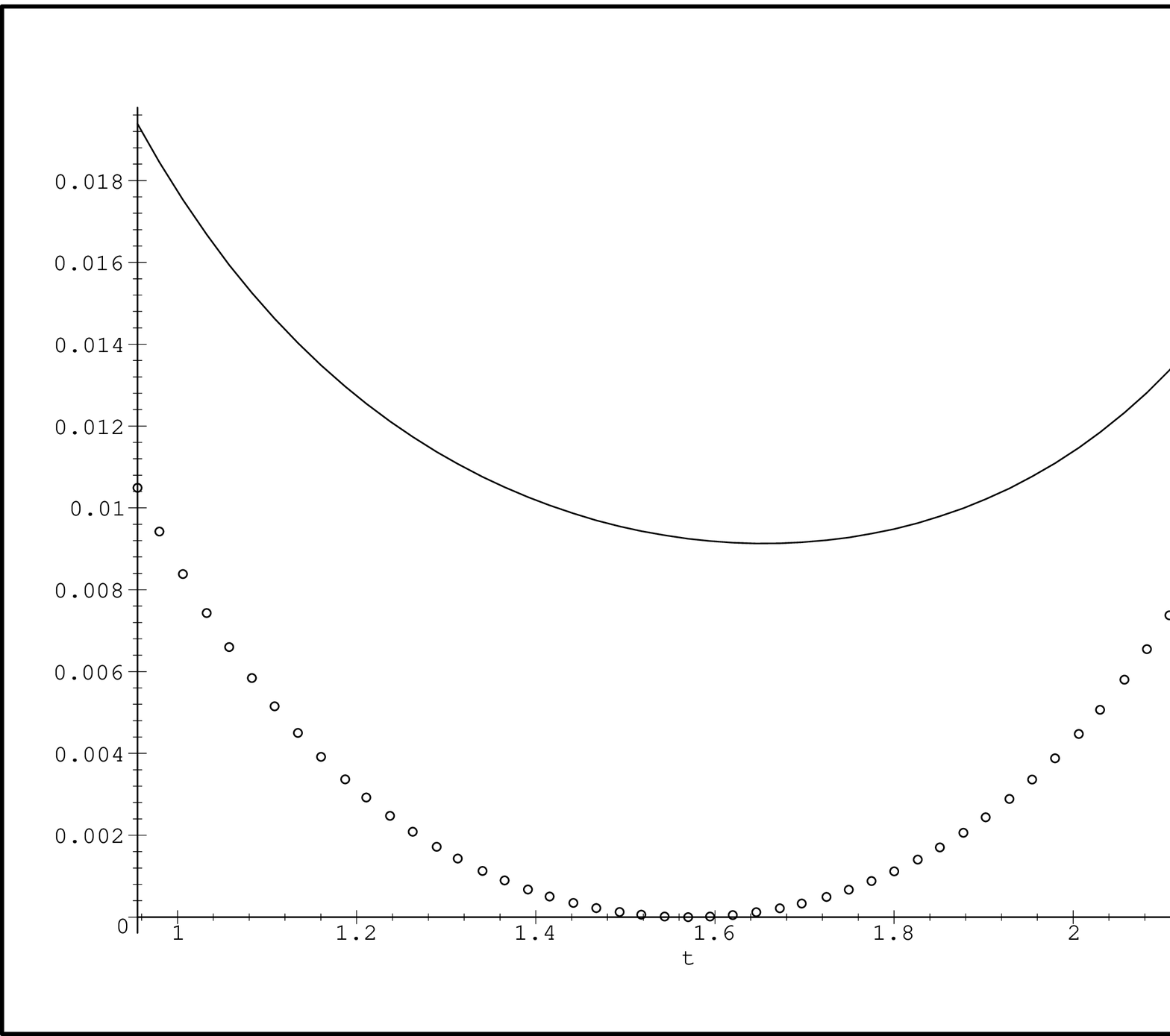}}}
\vskip .2 cm
\centerline{{\bf Fig.1}: The left mass insertion $\delta^{L}_{db}$ as function
of $ \theta $ (rad),}
\centerline { the
full line is the renormalised curve (see text).}
\vskip .5 cm

 From $(6.5), (6.6), (6.10)$ and $(4.6)$, one obtains the predictions for
the matrices $(4.7),(4.8)$ in this particular model. It is then possible
 to vary $\theta$ and plot
$\delta^{D}_{L,i,j}$ as  functions of $\theta$ (see figs. 1,2 and 3
where $\theta$ is measured in radians). As a
rule, effects of renormalisation are quite severe. On each plot we have
depicted two curves, one without renormalisation but for the gluino
contribution $ 7M^2_0 $ to the average squark $ {\rm \bar mass}^2 $, compared
to the
renormalised one.  As already discussed in the introduction,
 the net effect of renormalisation is to reduce the value of mass
insertions as the squark masses get a large common contribution
proportional to the gluino mass $[12,13,14]$. Otherwise, the
effect of renormalisation is small for the (ds) insertion whereas it
is large for the other two. Notice that the chosen range for $\theta$ is such
that mass insertions are within the phenomenological bounds
discussed in $ [8,10,11] $.

\vskip -3.0 cm
\epsfysize=12.cm{\centerline{\epsfbox{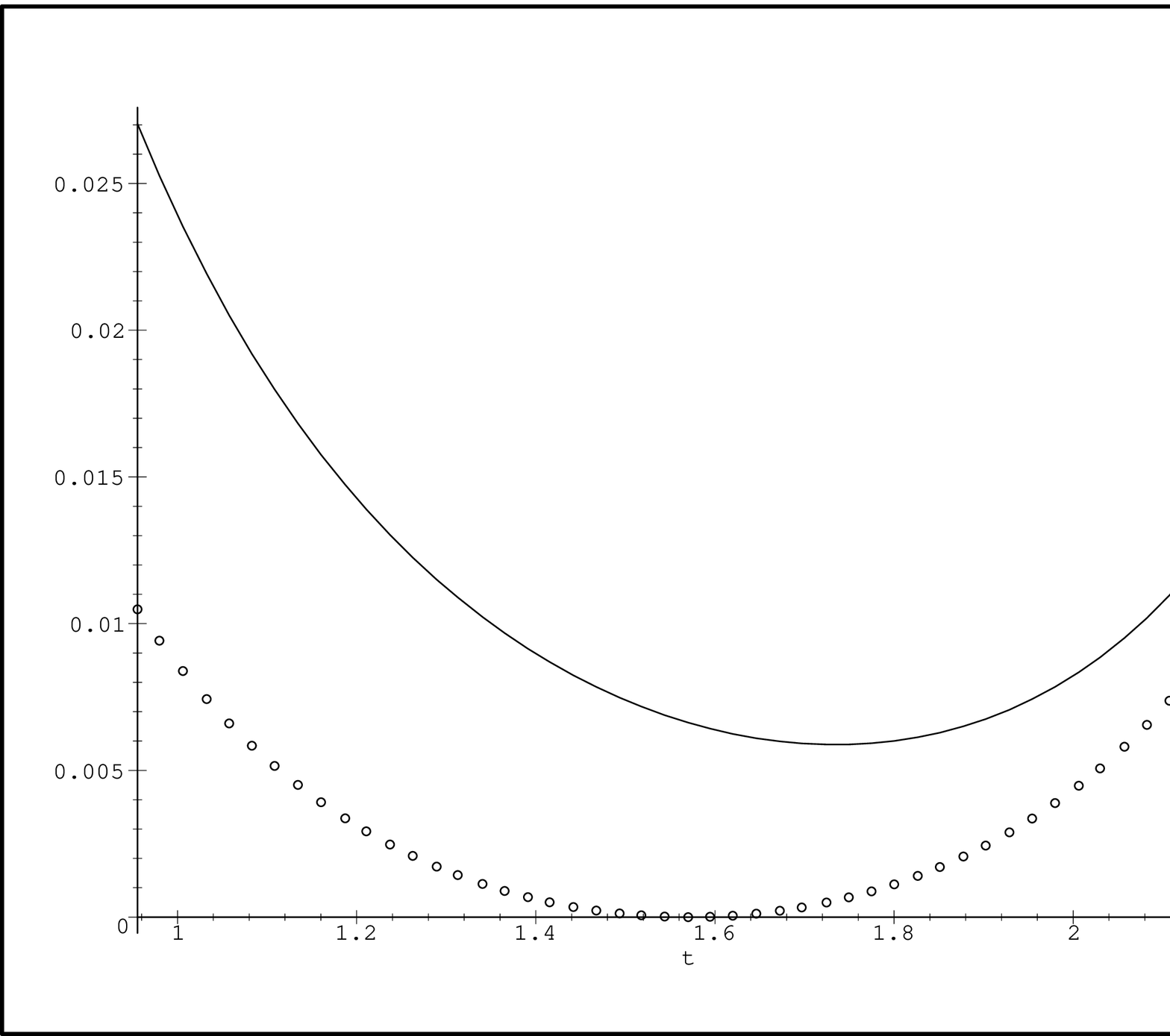}}}
\vskip .2 cm
\centerline{{\bf Fig. 2} The mass insertion $\delta^{L}_{sb}$ as function
of $ \theta $ (rad),}
\centerline { the
full line is the renormalised curve (see text).}
\vskip .5 cm

Because of the recent measurements of the $ b \longrightarrow s \gamma $
transition, let us consider the helicity flipping analogue of $ (4.7), (4.8)$.
They are given by the flavour mixing elements in the analytic sector of the
${\rm mass}^2$ matrix. Let us define,
$$ \tilde \delta ^D_{LR} = {{K(B^D_Rm_D+m_DB^D_L)K_{\dagger}} \over
{\bar m^2}} . \eqno (6.12) $$

and use $(6.5)$, $(6.6)$ and $(4.6)$ to obtain the following
numerical expressions:
$$ \eqalignno{ \left(\tilde \delta^ D_{LR} \right)_{sd} & =
-(.05 \cot \theta + 0.9){{m_s}\over {M_0}} \cr
\left(\tilde \delta^ D_{LR} \right)_{db}  & = - .04 \cot \theta {{m_b} \over
{M_0}} \cr
\left(\tilde \delta^ D_{LR} \right)_{sb} & = -(.06 \cot \theta +.015)
{{m_b} \over {M_0}} . & (6.13) \cr} $$

which are all below the experimental bounds.

\vskip -3.0 cm
\epsfysize=12.cm{\centerline{\epsfbox{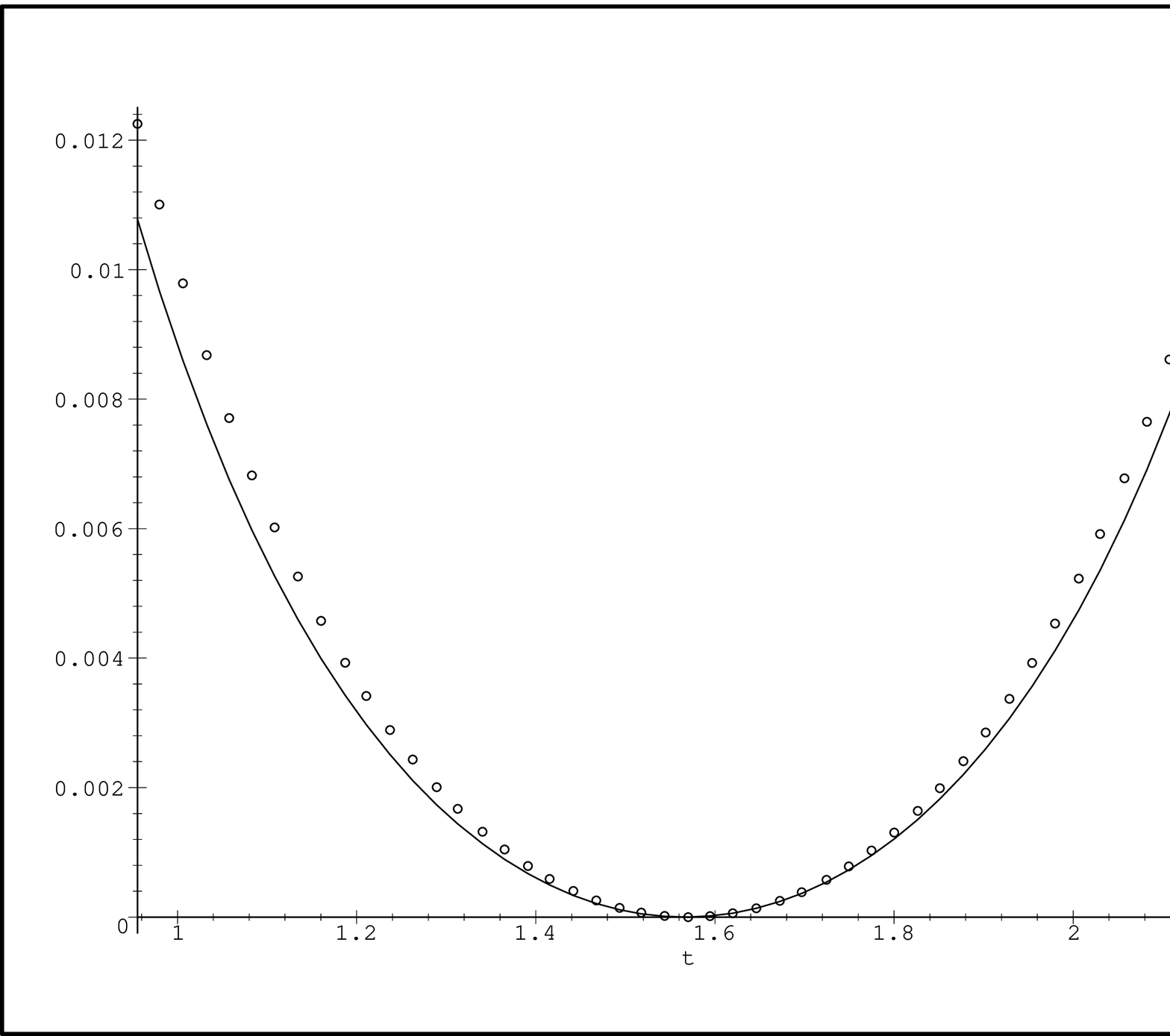}}}
\vskip .2 cm
\centerline{{\bf Fig. 3} The mass insertion $\delta^{L}_{ds}$ as function
of $ \theta $ (rad),}
\centerline { the
full line is the renormalised curve (see text).}
\vskip .5 cm

Since we are interested  in renormalisation effects,
we have concentrated on the quark sector. Indeed, quantum corrections are much
smaller in the
leptonic sector where the large QCD and top couplings do not contribute
(quantum effects can be bigger in the large $ \tan \beta $ scenario).
However the experimental FCNC restrictions are much stronger because of the
limits on $ \mu \longrightarrow e \gamma $ and similar processes.
 Let us generalise the mixing matrix U in $ (6.2)$ to
the leptons, choose the lepton modular weights to be those assigned to quarks
in the \lq \lq same family\rq \rq .
Then we can  estimate the analogues of $(4.8)$ and $(6.12)$ in the
lepton sector. By imposing the experimental bounds for
$\mu \longrightarrow
e \gamma $ on the resulting expressions for
$(\delta_L)_{e\mu}$ ,$(\delta_R)_{e\mu}$ and $(\delta_{LR})_{e\mu}$, one gets
a strong constraint on the Goldstino angle, $ \tan ^2 > 10 $. With this
restriction on $\theta$, the values of $(\delta^D_L)_{ij}$ in the figures
lie well below their phenomenological limits.

In summary, we have considered a simple model to estimate FCNC effects
including the RGE running, with maximal family mixing at the
unification scale. The flavour dependent supergravity couplings
is due to the moduli fields, represented in our example by a
single superfield, $T$. Thus FCNC effects are proportional to $\cot \theta$
( or $\cot ^2 \theta $), which measures the supersymmetry breaking component
along the flavour dependent direction. The resulting FCNC effects are
small for two reasons: (i) The  universal gluino contribution
to the squark masses is very large, a model-independent feature. (ii) The
contributions of the moduli sector to the scalar ${\rm mass}^2$ are
proportional to $ (n_i+1) \cot ^2 \theta $ which are basically non-positive
if the modular weights $ n_i \le -1$ as usually found in orbifold models:
this fact sets a physical upper limit on the supersymmetry breaking
component along the moduli direction.
Expressed as a bound on $\cot \theta$ it restricts the FCNC effects to
the relatively small values in the figures, even in the maximal mixing
model discussed here.

This negative contribution to the squark  ${\rm mass}^2$ may be considered
as an artefact of the model, or more generally, of flavour dependence
induced by the moduli couplings. Even if this is an interesting natural
mechanism to generate non-universal supergravity couplings, it is
certainly possible to contemplate other possibilities. However, unless
 quarks and leptons exhibit very different patterns, there is another general
constraint on FCNC effects. Indeed, whatever mechanism generates flavour
dependence, the mixing effects in the lepton sector are severely restricted
by experiments. Hence supersymmetry breaking auxiliary fields with flavour
dependent couplings are bounded, as in the example here above, and quark
flavour
changing effects are expected to be well below the experimental limit.
But, a priori, quarks and leptons could  couple to supergravity
in quite different ways.

\vskip 17pt
\noindent {\bf 7. CONCLUSION}
\vskip 12pt
Within the supergravity framework, most of the unitary transformations needed
to diagonalise the quark masses acquire a physical meaning. The RGE
are then needed to make the connection with  low energy flavour mixing
phenomena. The systematic
study presented here, shows how the quark mass hierarchies helps
simplifying results. This is so even if the tree-level
mixing is large. The effects of the running down to low energies
of the squark mass matrices is sizeable in many respects, and
the factorisation of the A-terms into $B_L$ and $B_R$ is a useful
tool.  Nevertheless, some uncertainties remain due to the
dependence on the unknown analogue of the CKM matrix for right-handed
quarks.

 Approximate discrete symmetries could be invoked
to justify that
the large flavour mixing in the Yukawa couplings would disfavour
non-universality in the Kahler potential. An elegant analysis $[22]$
shows how  ad hoc broken continuous symmetries could naturally
control non-universality patterns in the squark masses
and prevent FCNC problems.
It would be relevant to study how natural  these symmetries are in the
context of supergravity.

\vfill\eject

\centerline{{\bf REFERENCES}}
\vskip 24pt
\item{$\lbrack$1$\rbrack$} S.L. Glashow, J. Iliopoulos and L. Maiani, Phys.
Rev. {\bf D2} (1970)
1285.

\item{$\lbrack$2$\rbrack$} J. Ellis and D. Nanopoulos, Phys. Lett. {\bf 110B}
(1982) 44.
\item{\nobreak\ \nobreak\ \nobreak\ } R. Barbieri and R. Gatto, Phys. Lett.
{\bf 110B} (1982) 211.
\item{\nobreak\ \nobreak\ \nobreak\ } T. Inami and C.S. Lim, Nucl. Phys. {\bf
B207} (1982) 533.
\item{\nobreak\ \nobreak\ \nobreak\ } B.A. Campbell, Phys. Rev. {\bf D28}
(1983) 209.
\item{\nobreak\ \nobreak\ \nobreak\ } M.J. Duncan, Nucl. Phys. {\bf B221}
(1983) 285.
\item{\nobreak\ \nobreak\ \nobreak\ } E. Franco and M. Mangano, Phys. Lett.
{\bf 135B} (1984) 445.

\item{$\lbrack$3$\rbrack$} J.F. Donoghue, H.P. Nilles and D. Wyler, Phys.
Lett. {\bf B128} (1983)
55.

\item{$\lbrack$4$\rbrack$} A. Bouquet, J. Kaplan and C.A. Savoy, Phys. Lett.
{\bf B148} (1984) 69.

\item{$\lbrack$5$\rbrack$} M. Dugan, B. Grinstein and L.J. Hall, Nucl. Phys.
{\bf B255} (1985) 413.

\item{$\lbrack$6$\rbrack$} L.J. Hall, V.A. Kostelecky and S. Raby, Nucl. Phys.
{\bf B267} (1986)
415.

\item{$\lbrack$7$\rbrack$} S. Bertolini, F. Borzumati and A. Masiero, Phys.
Lett. {\bf B194}
(1987); {\bf B192} (1987) 437.

\item{$\lbrack$8$\rbrack$} F. Gabbiani and A. Masiero, Nucl. Phys. {\bf B322}
(1989) 235.

\item{$\lbrack$9$\rbrack$} S. Bertolini, et al., Nucl. Phys. {\bf B353} (1991)
591.

\item{$\lbrack$10$\rbrack$} J.S. Hagelin, S. Kelley and T. Tanaka, Nucl. Phys.
{\bf B415} (1994)
293.

\item{$\lbrack$11$\rbrack$} Y. Nir and N. Seiberg, Phys. Lett. {\bf B309}
(1993) 337.

\item{$\lbrack$12$\rbrack$} A. Brignole, L. Ibanez and C. Munoz, Nucl. Phys.
{\bf B420} (1994) 125.

\item{$\lbrack$13$\rbrack$} D. Choudhury et al., Max Planck preprint
MPI-PHT-94-51 (1994).
\item{\nobreak\ \nobreak\ \nobreak\ \nobreak\ } J. Louis and Y. Nir, M\"unich
preprint LMU-TPW-94-17 (1994).

\item{$\lbrack$14$\rbrack$} P. Brax and M. Chemtob, Saclay preprint T94/128.

\item{$\lbrack$15$\rbrack$} For a review see, e.g., H.P. Nilles, Phys. Rep.
{\bf 110} (1984) 1.

\item{$\lbrack$16$\rbrack$} For a review see, e.g., M. Green, J.S. Schwarz and
E. Witten,
\lq\lq Superstring Theory\rq\rq , Cambridge University Press, 1987.

\item{$\lbrack$17$\rbrack$} L. Dixon, V. Klapunovsky and J. Louis, Nucl. Phys.
{\bf B329} (1990)
27; {\bf B355} (1991) 649.
\item{\nobreak\ \nobreak\ \nobreak\ \nobreak\ } V. Klapunovsky and J. Louis,
Phys. Lett. {\bf B306} (1993) 269.

\item{$\lbrack$18$\rbrack$} L. Ibanez and D. Lust, Nucl. Phys. {\bf B382}
(1992) 305.
\item{\nobreak\ \nobreak\ \nobreak\ \nobreak\ } B. de Carlos, J.A. Casas and
C. Munoz, Phys. Lett. {\bf B299} (1993)
234.

\item{$\lbrack$19$\rbrack$} J.P. Derendinger, S. Ferrara, C. Kounnas and F.
Zwirner, Nucl.
Phys. {\bf B377} (1992) 145.

\item{$\lbrack$20$\rbrack$} R. Barbieri, J. Louis and M. Moretti, Phys. Lett.
{\bf B312} (1993)
451.

\item{$\lbrack$21$\rbrack$} L. Ibanez and G.G. Ross, Phys. Lett.
{\bf B332} (1994) 100.

\item{$\lbrack$22$\rbrack$} M. Leurer, Y. Nir and N. Seiberg, Nucl. Phys.{\bf\
B420} (1994) 468.

\item{$\lbrack$23$\rbrack$} R. Barbieri, S. Ferrara and C.A. Savoy, Phys.
Lett. {\bf B119} (1982)
343.
\item{\nobreak\ \nobreak\ \nobreak\ \nobreak\ } R. Arnowitt, A. Chamseddine
and P. Nath, Phys. Rev. Lett. {\bf B49}
(1982) 970.
\item{\nobreak\ \nobreak\ \nobreak\ \nobreak\ } L. Hall,
 J. Lykken and S. Weinberg, Phys. Rev. {\bf D27} (1983)
2359.

\item{$\lbrack$24$\rbrack$} A. Soni and H. Weldon, Phys. Lett. {\bf B126}
(1983) 215.

\item{$\lbrack$25$\rbrack$} M. Olechowski and S. Pokorski, Phys. Lett.
{\bf B257} (1991) 388.

\item{$\lbrack$26$\rbrack$} J. Ellis, S. Ferrara and D.V. Nanpoulos, Phys.
Lett. {\bf B114} (1982)
231.
\item{\nobreak\ \nobreak\ \nobreak\ \nobreak\ } J. Polchinski and M.B. Wise,
Phys. Lett. {\bf B125} (1983) 393.
\item{\nobreak\ \nobreak\ \nobreak\ \nobreak\ } W. Buchmuller and D. Wyler,
Phys. Lett. {\bf B121} (1983) 321.
\item{\nobreak\ \nobreak\ \nobreak\ \nobreak\ } M. Dugan, B. Grinstein and L.
Hall, Nucl. Phys. {\bf B255} (1985) 413.

\item{$\lbrack$27$\rbrack$} A. Bouquet, J. Kaplan and C.A. Savoy, in \lq\lq
Third CSIC Workshop on
SUSY and Grand Unification\rq\rq , edited by J. Leon et al. (World Scientific,
Singapore, 1986) p.373.

\item{$\lbrack$28$\rbrack$} L.E. Ibanez and C. Lopez, Nucl. Phys. {\bf B233}
(1984) 511.
\item{\nobreak\ \nobreak\ \nobreak\ \nobreak\ } A. Bouquet, J. Kaplan and C.A.
Savoy, Nucl. Phys. {\bf B262} (1985)
299.

\end